\documentclass[preprint]{aastex}
\input epsf

\def\etal{{et al.~}}
\shorttitle{GCs in NGC 7457}
\shortauthors{Chomiuk \etal}

\begin{document}

\title{The Peculiar Globular Cluster System of the S0 Galaxy NGC 7457}

\author{Laura Chomiuk\altaffilmark{1}, Jay Strader\altaffilmark{2,3}, \& Jean P. Brodie\altaffilmark{4}}
\email{chomiuk@astro.wisc.edu, jstrader@cfa.harvard.edu, brodie@ucolick.org}

\altaffiltext{1}{University of Wisconsin--Madison, Madison, WI 53706}
\altaffiltext{2}{Harvard-Smithsonian Center for Astrophysics, Cambridge, MA 02138}
\altaffiltext{3}{Hubble Fellow}
\altaffiltext{4}{UCO/Lick Observatory, University of California, Santa Cruz, CA 95064}

\begin{abstract}

We present \emph{HST} photometry and Keck spectroscopy of globular clusters (GCs) in the nearby S0 galaxy NGC
7457. The $V-I$ color-magnitude diagram of GCs lacks the clear bimodality present in most early-type galaxies;
there may be a significant population of intermediate-color objects. Of 13 spectroscopically-observed GCs, two
are unusually metal-rich and feature bright [\ion{O}{3}] emission lines. We conclude that one probably hosts a
planetary nebula and the other a supernova remnant. Such emission line objects should be more common in an
intermediate-age stellar population than in an old one. We therefore suggest that, in addition to the typical old
metal-rich and old metal-poor GC subpopulations, there may be a third subpopulation of intermediate age. Such a
subpopulation may have been formed $\sim 2-3$ Gyr ago, in the same star-forming event that dominates the stellar
population of the center of the galaxy.

\end{abstract}

\section{Introduction}

Extragalactic globular clusters (GCs) provide a unique window onto the stellar populations of nearby galaxies, as
they trace and simplify star formation episodes in nearly all galaxies from dwarfs to giant ellipticals. If a
subpopulation of GCs is found to have a given age and metallicity, one can extrapolate that there is a
significant component of the parent galaxy's stellar population with the same characteristics (Brodie \& Strader
2006, Harris 2001, and references therein). In this way, GCs are valuable fossils which allow us to dissect the
complex star formation histories of galaxies in our local universe.

S0 galaxies stand at a junction in the Hubble tuning fork, and they may symbolize a transition in the
dominant processes of galaxy evolution. As summarized by Kormendy \& Kennicutt (2004), in the past, galaxy
evolution was dominated by the violent processes of hierarchical merging. In the future, the main processes
affecting galaxies will be more gradual, involving the detailed internal dynamics of galaxies themselves.  
Today, the relative importance of secular and merging processes appears to be a function of Hubble type, as the
bulges of late-type spirals tend to be formed by gas and stars funneled along disk asymmetries like bars, while
ellipticals are thought to have been formed by mergers and violent relaxation. However, the relative importance
of these mechanisms is still unclear in early-type spirals and lenticulars. The bulges of these galaxies appear
to be mostly ``classical" (formed by mergers), but secular evolution certainly plays a role in many systems.

The SA0 galaxy NGC 7457, at first glance, looks like a perfect example of a normal lenticular, with a classical
de Vaucouleurs bulge and an exponential disk (Kormendy 1977). Basic data for this galaxy are listed in Table 1.
No dust and very little cold gas have been detected in NGC 7457 (Peletier \etal 1999, Welch \& Sage 2003, Sage 
\& Welch 2006). It is one of the nearest S0 galaxies, at a distance of 13 Mpc (Cappellari \etal 2006). However, on 
closer inspection, NGC 7457 has shown several peculiarities in its dynamics and stellar populations, which have 
made it the center of much debate. Kormendy \& Illingworth (1983)  observed that the bulge of NGC 7457 has an 
unusually low velocity dispersion for its luminosity. This suggested that the bulge has the more ordered dynamics 
of a disk, and is predominantly supported by rotation. However, Dressler \& Sandage (1983) noticed that the bulge 
also appeared to have unusually low circular velocities. This puzzling contradiction has been explained by SAURON 
data, which show that NGC 7457 actually has a counter-rotating core (Sil'chenko \etal 2002, Emsellem \etal 2004). 
A counter-rotating core is a possible consequence of a merger. Additionally, the core has a distinct stellar population 
from the bulge. Sil'chenko \etal (2002) found that the bulge has an intermediate luminosity-weighted age of 5-7 Gyr, 
while the compact nucleus (radius of 1.5\arcsec) contains a younger (2-2.5 Gyr), chemically distinct stellar population.

GCs may be an interesting diagnostic of the relative significance to star formation of violent vs. secular
processes. If stars in a galaxy were formed from a gaseous merger, stars will have formed in a burst, and GCs
along with them. If stars form more quiescently, then GCs may not accompany them (Brodie \& Strader 2006).
Therefore, one of the goals of this study is to search for GCs of intermediate age to place constraints on the
bulge and nucleus formation in NGC 7457.

Several claims of unusual GC phenomena have been made for the NGC 7457 system by Chapelon \etal (1999). They
obtained ground-based B, V, and I photometry, but did not find the usual color bimodality in the GC population.  
This color bimodality is observed in the Milky Way and many other galaxies of all morphological types, and
generally represents two old stellar populations with different metallicities (see Brodie \& Strader 2006, and
references therein). In the Milky Way, these populations have [Fe/H] = $-1.5$ (for the blue peak), and [Fe/H] =
$-0.5$ (for the red peak). Chapelon \etal found a broad unimodal distribution centered at colors corresponding to
[Fe/H] = $-1.0$, with a width of $\sigma$([Fe/H]) = 0.6 dex. The authors claim that this width is too narrow to
be composed of two unresolved GC populations, and that the mean color is too red to accommodate a significant
population of blue (metal-poor) GCs. Provocative as these results are, the ground-based data have a small
sample size and large photometric errors. A solid understanding of the NGC 7457 GC system requires a larger
sample, more accurate photometry, and spectroscopy.

In Section 2, we discuss archival \emph{Hubble Space Telescope (HST)} photometry of the NGC 7457 GC system. In Section
3 is presented Keck/LRIS spectroscopy for 13 GCs. Section 4 provides an analysis of the radial velocities
obtained from the spectroscopy, while in Section 5 we present our techniques for constraining the spectroscopic
GCs' stellar populations.  In Sections 6 and 7 we discuss the properties of the stellar populations. In Section
8, we discuss two curious GCs with emission lines. Finally, in Section 9, we synthesize our results and draw
conclusions.

\section{HST Photometry}

\subsection{WFPC2 Data}

We downloaded archival \emph{HST}/WFPC2 images of the center of NGC 7457 taken in Cycle 4. These images have
total exposure times of 1250 sec in F814W and 1330 sec in F555W, and were pipeline-processed in the usual way
before retrieval. The individual images were registered and coadded to reject cosmic rays. Detection of GC
candidates was carried out on 20 $\times$ 20 pixel median-subtracted images. Because of the low density of GC
candidates, aperture photometry is sufficient. We measured magnitudes through apertures with 3 pixel radii
(corresponding to 0.3\arcsec\ for the WF chips and 0.14\arcsec\ for the PC chip). These were corrected to
standard $V$ and $I$ magnitudes using the updated equations of
A.~Dolphin\footnote{http://purcell.as.arizona.edu/wfpc2\_calib/} (see Dolphin 2000), except that no correction
for charge transfer efficiency was made due to the galaxy background.

Corrections from 3 to 5 pixel apertures for the WF chips were taken from Larsen \& Brodie (2000), and assume GC
effective radii of 3 pc and a galaxy distance of 10 Mpc. These are $-0.14$ mag in $V$ and $-0.166$ in $I$. As
noted by Larsen \& Brodie, these corrections will systematically underestimate the magnitudes of very large GCs,
but the colors are unaffected. An additional $-0.1$ mag was added to correct the magnitudes to a nominal infinite
aperture (Holtzman \etal~1995). Larsen \& Brodie (2000) did not calculate aperture corrections for the PC chip,
so we used the equations in Kundu \& Whitmore (2001), which give corrections of $-0.44$ mag in $V$ and $-0.49$
mag in $I$. These are from 3 pixels to infinity (thus including the Holtzman \etal~correction) and assume a
galaxy distance of 13 Mpc. Finally, the $V$ and $I$ magnitudes were corrected for extinction, using values of
$A_{V} = 0.172$ mag and $A_{I} = 0.101$ mag, from the maps of Schlegel \etal~(1998). All plotted and tabulated
magnitudes are extinction-corrected.

GC candidate sizes were estimated using the \emph{ishape} algorithm (Larsen 1999), which convolves subsampled
TinyTim model PSFs with King models of varying effective radii to find the best fit to the data. We used a fixed
model concentration $c = r_{tidal}/r_{core} = 30$. The resulting initial list of GC candidates was trimmed using
a color cut of $0.7 < V-I < 1.5$ and an upper size limit of 25 pc.

These criteria left us with a final list of 75 GC candidates. The associated $V$ vs.~$V-I$ color-magnitude
diagram is shown in Figure 1, and Figure 2 shows a histogram of the $V-I$ color distribution. It is difficult to
draw detailed conclusions from these plots due to the small number of objects and the moderately shallow nature
of the imaging. Nonetheless, it is clear that there is a broad distribution of GC colors, with perhaps two
distinct peaks: a dominant population of GCs around $V-I \sim 1.0$ and a smaller, redder set of objects. A
clearly bimodal distribution of GCs, like that seen in most early-type galaxies (Brodie \& Strader 2006), is not
obvious. We used the program Nmix (Richardson \& Green 1997; also see discussion in Strader \etal~2006) to
perform mixture modeling on the colors of GCs with $V < 23.5$; this analysis yields a strong preference for a
unimodal distribution peaked at $V-I = 1.02$.

These results are generally consistent with those of Chapelon \etal~(1999), who used ground-based imaging of NGC
7457 to claim a broad GC color distribution centered around $V-I = 1.04$. They suggested the existence of two
subpopulations of GCs, but with a smaller peak separation than is typical for massive early-type galaxies.

One plausible alternative scenario is that there is an underlying bimodal distribution of old metal-poor and
metal-rich GCs (at the typical metallicities) with an added younger component at intermediate colors. According
to the models of Maraston (2005), a single stellar population (SSP) with solar metallicity, an age of 1.5 Gyr,
and a Kroupa initial mass function should yield a color of $V-I = 1.02$; at lower metallicities, this color
corresponds to somewhat older ages. Such an age would be consistent with the spectroscopic age of the center of
the galaxy (Sil'chenko \etal 2002; see Section 1). However, if this younger component formed with a standard GC
mass function, the clusters ought to extend to brighter magnitudes than the old GCs (as seen, for example, in the
merger remnant NGC 1316; Goudfrooij \etal~2001). This behavior is not obvious in Figure 1, perhaps indicating
that there are only a small number of intermediate-age GCs, or that they did not form with the GC mass function
typical in luminous galaxies. Small number statistics could also play a role.

Even if there is a significant subpopulation of intermediate-age GCs, both of the old GC subpopulations are
probably still present in NGC 7457. It is unlikely that all of the red GCs are of intermediate or younger age,
given that they extend to rather red colors ($V-I > 1.2$) difficult to produce except at old ages. In addition,
no luminous galaxy has been convincingly shown to lack metal-poor GCs. If we model the GC colors with the
restriction that there are three subpopulations, the color peaks lie at $V-I$ = 0.92, 1.03, and 1.16 (with
relative contributions of 33\%, 41\%, and 25\%). The extreme red and blue peaks are consistent within $\sim$0.02
mag of the GC color--galaxy luminosity relations for the old metal-rich and metal-poor subpopulations (Strader
\etal 2004). We thus consider a three-component GC system to be the most likely explanation, though by no means
certain. Given the limited spatial coverage of the WFPC2 data, the small number of objects, and the uncertain
mass function of the hypothesized intermediate-age GCs, the percentages above should be considered very
preliminary, but they do indicate that the younger objects might represent a substantial fraction of the overall
GC system.

\subsection {ACS/HRC Data}

As this manuscript was nearing submission, ACS/HRC UV imaging of the central regions of NGC 7457 became publicly
available. This imaging consisted of 2552 sec in F250W, 1912 sec in F330W, and 432 sec in F555W. Since this
imaging partially overlaps the older WFPC2 imaging discussed above, we decided to analyze it to increase our
wavelength baseline, potentially increasing our ability to detect intermediate-age GCs. While the F555W image is
useful for morphological classification of GC candidates, it is not substantially deeper than the WFPC2 image and
so we did not perform photometry on it.

The analysis process was similar to that used for the WFPC2 images. We searched for GC candidates on
median-subtracted, pipeline-processed images. These catalogs were cross-matched with the WFPC2 candidate list.
For the F250W image there were only two matches, and both objects appeared to be very faint, so we do not discuss
the F250W data further. However, there were eleven clear detections on the F330W image. Two of these were very
close to each other and to the center of the galaxy; the high background makes accurate photometry impossible.
They are located at a projected distance of 1.4\arcsec\ ($\sim 88$ pc) from the galaxy center, but are separated
from each other by only $\sim 0.2$\arcsec\ (13 pc). Assuming these are true GCs, it is very unlikely that they
form a bound pair; the merging timescale would be only a few $\times\ 10^7$ yr (Minniti \etal~2004; Sugimoto \&
Makino 1989).

For the other nine candidates, we performed aperture photometry in F330W with a 5-pixel aperture; this is a
compromise between the small pixel scale of the HRC (compared to WPFC2 and the intrinsic cluster sizes of $\sim$
0.03 - 0.05\arcsec) and the larger instrumental PSF in the ultraviolet. Since the GCs are resolved, the
``standard" encircled energy values cannot be used for aperture corrections; instead, these were estimated by
convolving a GC of typical size (2.5 pc at 13 Mpc) with a TinyTim ACS/HRC PSF using \emph{ishape} (Larsen 1999).  
The correction from our 5 pixel aperture to one of 0.5\arcsec\ is $-0.257$ mag. We then used the Sirianni
\etal~(2005) value of $-0.117$ mag to correct from a 0.5\arcsec\ to a nominally infinite aperture. The final
correction is for foreground reddening; using the Schlegel \etal~(1998) value of $E(B-V)=0.052$ and the
conversion factor in Sirianni \etal~(2005) yields $A_{F330} = 0.26$. The extinction and aperture corrections
total $-0.634$.

The dereddened $F330W-V$ colors of these nine candidate GCs are given in Table 2. In Figure 3 we show a $F330W-V$
vs.~$V-I$ color-color plot of the clusters. Overplotted are model isochrones from Girardi \etal~(2002) for two
ages: 2 Gyr and 11 Gyr, with metallicities ranging from [$Z$/H] = $-1.7$ to solar. The two objects that are
spectroscopically-confirmed members are marked separately. Seven of the objects are consistent with the model
isochrones, with four of the seven consistent with either old ages and low metallicities or with younger ages
(none of these four have spectra). The other three objects are too red in both $F330W-V$ and $V-I$ to be of
intermediate age. The single object marked with a circle (ID L1 in Table 2) is the most luminous in the sample,
with $V = 20.5$, and is also unusually large, with an effective radius of $\sim 20$ pc. Its UV-optical colors are
marginally more consistent with an intermediate age than the other candidates. The luminosity and size of the
cluster are reminiscent of NGC 2419 in the Milky Way, which has been argued to be the stripped core of a dwarf
galaxy (van den Bergh \& Mackey 2004). Spectroscopy of this object would be desirable to confirm membership in
NGC 7457 and to assess its age and metallicity.

Two candidates do not fall on the isochrones. One is the confirmed cluster GC11. Morphologically this object is
quite odd; it is notably asymmetric, with a ``tail"-like structure to the NE, and would likely be discarded as a
background galaxy in a typical study. Given its unusual colors and morphology, we suggest that this is a true GC
superimposed on a background object. The other object that is not on the isochrones (UV4) looks compact and
symmetric in all of the ACS/HRC images. It is probably a compact background galaxy, though we cannot exclude an
unidentified error with the photometry.

\section{Spectroscopic Observations and Data Reduction}

We observed the NGC 7457 GC system with Keck I and the Low-Resolution Imaging Spectrometer (LRIS; Oke \etal 1995)  
in multi-slit mode on October 21-22, 2003. A total of 15 exposures were taken, each of 1800 seconds duration, for
an equivalent exposure time of 450 minutes. We also obtained long-slit spectra of one flux standard star,
BD+28$^\circ$ 4211 from Massey \& Gronwall (1990), and four Lick/IDS standard stars, HR 2153, HR 2459, HR 7429,
and HR 8165 (Worthey \etal 1994). All data were obtained through a 600 line mm$^{-1}$ grism blazed at 4000 \AA\
on the blue side, and an 831 line mm$^{-1}$ grism blazed at 8200 \AA\ on the red side, with a dichroic splitting
the incoming light at 6800 \AA. Slits were 0.8\arcsec\ in width. On the blue side, this provides a reciprocal
dispersion of 0.63 \AA\ pixel$^{-1}$ and a spectral resolution of 2.8 \AA. The wavelength range of a slit in the
center of the mask is 3300-5880 \AA, but this range can vary by several hundred angstroms depending on the
positions of individual slitlets on the mask. The red side of our spectra covers the wavelength range 7100-9100
\AA\ on average, in order to include the \ion{Ca}{2} triplet. Unfortunately, the fringing proved too severe for a proper
reduction of the red spectra; here we present only the blue spectra.

Our data were reduced using the standard image and spectral reduction tools in IRAF. The raw images were
bias-subtracted and flat-fielded using a normalized dome flat. The intensity of the quartz lamp varied strongly
with wavelength, so the large-scale spectral response was removed from the dome flats before they were applied.  
Also, the quartz lamp provided very little signal in the far blue, so we were not able to flat-field at all at
the shortest wavelengths.

We then extracted our spectra using \emph{apall}. We wavelength-calibrated the spectra using HgNeArCdZn
comparison arc lamps, and applied small zero-point corrections of a few angstroms by aligning the bright
\ion{O}{1} skyline to its proper wavelength of 5577.34 \AA. At this point, we combined the 15 spectra from the
separate exposures by averaging them together with sigma-clipping. The quality of these spectra are estimated in
Table 3, where the signal-to-noise ratio (S/N) per \AA\ around H$\beta$ is listed for the spectra. We
flux-calibrated our averaged spectra using the flux standard BD+28$^\circ$ 4211.

We chose GCs for spectroscopy primarily from the WFPC2 imaging described in Section 2.1. Given the suggestion of
Sil'chenko \etal (2002) that NGC 7457 may be an intermediate-age merger remnant, one of our primary goals was the
identification of intermediate-age GC candidates. Thus red GCs were preferentially selected for inclusion in the
slit-mask, though a number of blue objects were included as well to fill in the mask. Slits were placed on 17
WFPC2 objects. From these, we obtained good spectra for 13 GCs and the nucleus of NGC 7457 itself. The other
three slits had various issues: one contained a GC, but was so near the nucleus of NGC 7457 that the background
was very high, and it proved impossible to subtract the galaxy's spectrum properly. Another slit also contained a
GC, but the data were vignetted by a bracket in the spectrograph. Finally, one of the slits contained an object
with a velocity of $-412$ km s$^{-1}$, most likely a Galactic star. Additional GC candidates in the outer parts
of the galaxy were selected from ground-based imaging; unfortunately, our spectroscopy indicates that all of
these extra candidates are contaminants (7 stars and 2 background galaxies).

Basic information about the 13 GCs in our sample can be found in Table 3. The radial velocities and S/N values
were measured on our Keck LRIS spectra. Figure 4 shows two such spectra, for a prototypical metal-poor GC, namely
GC9, and a metal-rich GC, GC5. The reader should also be aware of two interesting outliers, GC6 and GC7, which
not only exhibit the usual absorption-line spectra, but also [\ion{O}{3}]$\lambda\lambda$4959,5007 emission
lines. The spectra for these two GCs can be seen in Figure 5 (along with GC5 for comparison), and will be
discussed in Section 8.

\section{Kinematics}

Radial velocities for our 13 GCs and the center of NGC 7457 are listed in Table 3. Using \emph{fxcor} in IRAF, we
cross-correlated our spectra with MILES SSP models (S{\'a}nchez-Bl{\'a}zquez \etal 2006; A. Vazdekis \etal 2008, in preparation) to
determine radial velocities and errors. We bootstrapped to estimate global kinematic quantities. We find a mean
radial velocity of 856$\pm$20 km s$^{-1}$. Given the rotation of the GC system (see below), and the asymmetric spatial distribution
 of our GC sample, this value is consistent with our measured radial velocity of the center of NGC 7457, 789$\pm$15 km s$^{-1}$. The 
 velocity dispersion of the 13 GCs is 69$\pm12$ km s$^{-1}$.

Figure 6 shows the spatial distribution of the GCs and their radial velocities superimposed on a representative
isophote of the galaxy. The general impression is that the GC system is rotating around the minor axis of the
galaxy. Fitting a simple solid-body rotation curve to the weighted velocities gives a best-fit axis of $\sim
160^{\circ}$, generally consistent with the photometric position angle of $\sim 130^{\circ}$ for the galaxy
itself. The GCs are also rotating in the same sense (with the southern objects blueshifted) as the galaxy
(Emsellem \etal~2004).  Of course, if there is a substantial degree of rotation in the GC system, the velocity
dispersion quoted above should be taken only as an upper limit to the pressure support for the kinematics of the
GCs.

Given the small number of GCs and their asymmetric distribution, we see little value in detailed analysis of the
kinematics. The system appears more rotationally supported than most early-type galaxies---the notable exception
being the Virgo dwarf elliptical VCC 1087 (Beasley \etal~2006). NGC 7457 may provide additional evidence that
disky populations of GCs are common in lower-mass galaxies.

\section{Techniques for Fitting SSP Models to GC Spectra}

\subsection{Lick/IDS Index Measurements}

The Lick/IDS system of absorption-line indices provides a standardized framework in which one can derive
metallicities, ages, and abundance ratios for stellar populations. The Lick/IDS system was originally created by
Burstein \etal (1984), and has been continuously revised through the years (Worthey \etal 1994; Trager \etal
1998). Considerable attention has also been paid to modeling the Lick/IDS indices and their variations with
stellar age and metallicity (Bruzual \& Charlot 2003; Maraston \& Thomas 2000; Maraston, Greggio, \& Thomas 2001;  
Worthey \etal 1994; Fritze-v. Alvensleben \& Burkert 1995). Lick/IDS indices have been very successful at lifting
the age-metallicity degeneracy, and can distinguish, for example, intermediate-age (3-4 Gyr) metal-rich GCs from
their much older counterparts, which would be indistinguishable from optical photometry alone (Strader \etal
2003b).

In order to match our spectra to the standardized Lick/IDS system, we smoothed our flux-calibrated spectra with a
wavelength-dependent Gaussian kernel as prescribed in Worthey \& Ottaviani (1997) and for the reasons described
in Larsen \& Brodie (2002) and Strader \etal (2003b). All indices and errors were measured using the C++ program
\emph{Indexf} (Cardiel \etal 1998). We measured indices on 13 GCs and on the inner $\sim$1.25$\arcsec$ of the
galaxy itself, for comparison. We also measured indices for four Lick standard stars, and calculated mean offsets
between our measurements and the Lick system. These small shifts were applied to the Lick indices of our 14
objects, listed in Table 4. Additionally, for each index, we determined the error in these offset by calculating
the standard deviation of the difference between our measurements and the standard Lick values. These standard
deviations were added in quadrature to the \emph{Indexf} errors to give the errors listed in Table 4.

\subsection{Line Diagnostic Plots}

Plotting Lick indices against one another allows us to visualize the distributions of GC ages, metallicities, and
abundance ratios. Such plots can be seen as Figures 7-10. Often, individual indices are combined to form a more
robust composite index---for example, $<$Fe$>$ (the average of Fe5270 and Fe5335). Model isochrone and
isometallicity lines are overplotted. In this study, we use the SSP models of Thomas \etal (2003) and Thomas
\etal (2004).

Although Mg$b$ is commonly used as the index sensitive to $\alpha$ elements, in this study we will use Mg2, as it
appears that unknown systematics are affecting our Mg$b$ measurements. Figure 7a plots the Mg2 index against the
Mg$b$ index, and at low metallicities it is clear that the Mg$b$ index sits below the model predictions. GCs from
published data sets (NGC 4365 (Brodie \etal 2005); Fornax (Strader \etal 2003a), NGC 1407 (Cenarro \etal 2007))  
sit comfortably along this line, thus implying our data are faulty at Mg$b$. There is a sky line at
$\lambda$5202\AA\ in the Mg$b$ and Mg2 bandpasses; we tried masking this line and remeasuring the spectra, and
the index values changed very little. When we compare Mg2 with Mg1 in Figure 7b, the measurements agree well with
model predictions. When Mg2 is compared with other $\alpha$-element indicators, like CN2 and Ca 4227, it remains
consistent with the models. Thus, here and throughout this paper, we will make use of the Mg2 index while
rejecting the Mg$b$ measurements.

\subsection{Composite Metallicities}

We derived metallicity estimates for the individual NGC 7457 GCs using the composite technique described in
Strader \etal~(2007). Briefly, Galactic GCs are used to define second-order polynomial relations between the
strength of individual Lick indices and total metallicity. The estimates from the different indices are then
combined to yield a robust measure of the GC metallicity with an associated error. In this study, the Balmer
lines were excluded from our calculations due to emission-line contamination in two objects.

Table 3 lists the metallicity estimates for the 13 GCs and the center of NGC 7457 itself. These composite
metallicities are calibrated for old stellar populations with metallicities roughly within the range $-1.8 \la$
[m/H] $\la 0$; there may be systematic offsets for younger ages and very metal-rich or metal-poor objects.

\subsection{Multi-index Fits via $\chi^{2}$ Minimization}

We used an additional route to take advantage of the information provided by many indices---a $\chi^{2}$
minimization algorithm described by Proctor \etal (2004), which simultaneously finds best-fitting GC ages,
[Z/Fe], and [$\alpha$/Fe]. This algorithm uses a grid of models from Thomas \etal (2003, 2004)  to fit between 8
and 15 Lick indices.

The results of this analysis are presented in Table 5. Note that the multi-index fits were run twice for the two
GCs exhibiting emission lines: once including all indices, and a second time excluding Balmer indices (which are
probably suffering fill-in by emission lines). The best-fit GC parameters do not change significantly between the
two runs.

\section {Abundance Ratios}

Are old GCs in other galaxies $\alpha$-enhanced, like Milky Way globular clusters? We should be able to assess
[$\alpha$/Fe] for NGC 7457 GCs using Lick indices, plotting an $\alpha$-element index (usually involving Ca or
Mg)  against an iron-sensitive index. However, this task turns out to be difficult in reality, suffering from
observational and theoretical uncertainties.

Figure 8 plots the Mg2 index against the mean of the Fe5270 and Fe5335 indices. SSP models for [$\alpha$/Fe] =
0.0 and [$\alpha$/Fe] = 0.3 (Thomas \etal 2003) are superimposed on the plot. The majority of GCs form a cloud
with large scatter at low metallicities. It is difficult to reach conclusions about their abundance ratios due to
the large observational errors, coupled with the convergence of the models at these low metallicities.

The [$\alpha$/Fe] values derived from the $\chi^{2}$-minimization multi-index fits, listed in Table 5, are
consistent with $\alpha$-element enhancement, although they have large errors. The weighted mean [$\alpha$/Fe]
for all 13 GCs is $0.3\pm0.1$. However, one of the metal-rich GCs with emission lines (GC7) appears more 
consistent with [$\alpha$/Fe] = 0 than [$\alpha$/Fe] = 0.3. GC5 appears to have a low [$\alpha$/Fe] in Figure 8, but 
its multi-line fit gives [$\alpha$/Fe] = $0.3\pm0.2$.

The center of NGC 7457 itself clearly has a near-solar abundance ratio. It sits firmly on the [$\alpha$/Fe] = 0.0
model track in Figure 8, and the multi-index fitter gives [$\alpha$/Fe] = $-0.03\pm0.03$.

\section {Ages and Metallicities}

We can break the age-metallicity degeneracy for our GCs by plotting an age-sensitive index (a Balmer line)
against a metallicity-sensitive index (such as Fe5270, Fe5335, or Mg2). In Figure 9, we plot H$\beta$ against
$<$Fe$>$.  Overplotted on our data are theoretical isochrone and isometallicity lines from Thomas \etal (2003)
which assume [$\alpha$/Fe] = 0.3. In this plot, all of our globular clusters appear old. They cluster around the
14 Gyr isochrone, but it is difficult to draw conclusions about their ages beyond the fact that they are $> 5$
Gyr old.  The multi-index fits listed in Table 5 also imply old ages for the GCs. The weighted mean age for the
11 GCs which do not exhibit emission lines is $11.2\pm1.6$ Gyr.

There is striking contrast between the globular clusters and the central region of NGC 7457 itself. The galaxy's
nucleus is much younger; in Figure 9, it appears to be dominated by stars aged $\sim$1.7 Gyr. However, keep in
mind that the $\alpha$-enhanced models are not appropriate for comparison with the galaxy spectrum, as the center
of NGC 7457 exhibits [$\alpha$/Fe]=0. The multi-index fit age-dates the center of NGC 7457 as $2.5\pm0.3$ Gyr
old, in agreement with the observations of Sil'chenko \etal (2002), who assigned an age of 2-2.5 Gyr.

The galactic nucleus is also significantly more metal-rich than even the most enhanced GC. Its multi-index fit
implies a super-solar metallicity of [Fe/H] = 0.18. This is more metal-rich than the solar metallicity found for
NGC 7457's nucleus by Sil'chenko \etal (2002), but this discrepancy is probably only a difference in models used;  
Sil'chenko \etal measure similar Lick index values, but they compare their data with models from Worthey (1994).

As seen in Figures 9 and 10, the GCs themselves occupy a broad spread in metallicity, as the GCs extend from low
metallicities to near-solar metallicities. No bimodality is apparent in these figures, and should not be expected
due to small sample size and inherent biases in our selection of spectroscopic candidates. The multi-index fits
imply [Fe/H] values in the range of $-1.6$ to $-0.3$, consistent with the composite metallicities.

The two GCs exhibiting emission lines (ELGCs), GC6 and GC7, are by far the most metal-rich GCs in our sample.  
Their metallicities are almost solar, with multi-index fits giving [Z/H] of $-0.1\pm0.1$ and $-0.2\pm0.2$. As
seen in Table 3, they are also the most metal-rich GCs according to their composite metallicities, and have
composite [m/H] quite similar to that of the center of NGC 7457. One must note the significant caveat that these
composite metallicities are calibrated using Galactic GCs, and so one should expect systematic deviations for
young ages and solar abundance ratios. When we take into account all the different methods for measuring
metallicity, it appears that GC6 and GC7 may have metallicities which are closer to the center of NGC 7457 than
to the 11 normal GCs.

Additionally, GC6 and GC7 are possible exceptions to the norm of old GC ages. Their Balmer absorption lines are
being filled in by emission, and therefore our Balmer index measurements should be seen as lower limits. They
appear to follow similar trends as the rest of the GCs in the H$\beta$ against $<$Fe$>$ plot, but we can also
assess their ages using a higher-order Balmer line, which should suffer less emission line fill-in.

Figure 10 plots H$\delta$F against $<$Fe$>$. Despite larger errors for H$\delta$F, the two ELGCs (seen as open
triangles) sit near the galactic center in this parameter space. All three objects have super-solar metallicity
in this plot. There are hints that the two ELGCs may have intermediate ages like the galactic center ($\sim$ 4--5
Gyr according to this plot).

A significant body of evidence therefore supports a scenario where the stellar populations of the two ELGCs
resemble the luminosity-weighted galaxy center more closely than they resemble the other normal GCs in the NGC
7457 system. However, the multi-index fits to GC6 and GC7 age-date them as old (11-15 Gyr), both when we include
and exclude the Balmer indices.  Of course, the metal-index Lick indices are primarily sensitive to metallicity,
but they retain some sensitivity to age. To our knowledge, deriving ages solely from metal lines has not been
attempted for objects known to be of intermediate age, so it is unclear how significant the multi-index derived
old ages for GC6 and GC7 are.

\section{Two Emission Line Objects in GCs}

\subsection{Association between GCs and the Emission Line Sources}

Both ELGCs are part of the NGC 7457 system. The radial velocities of GC6
(818$\pm$11 km s$^{-1}$) and GC7 (822$\pm$10 km s$^{-1}$) are consistent with the radial velocity of the
galaxy itself (789$\pm$15 km s$^{-1}$), confirming these objects as members of the NGC 7457 system. Additionally,
GC6 and GC7 have very similar radial velocities to one other.

The sources of the emission lines appear kinematically coupled to the clusters. In the case of GC6,
the [\ion{O}{3}]$\lambda\lambda$4959,5007 lines have a radial velocity of 857 km s$^{-1}$, while the stellar
absorption associated with this cluster has a radial velocity of 818$\pm$11 km s$^{-1}$. The [\ion{O}{3}] lines
in GC7 have a radial velocity of 837 km s$^{-1}$, in comparison with the stellar spectrum of the cluster receding
at 822$\pm$10 km s$^{-1}$. Our measurements of the emission line radial velocities are only based on two lines,
and therefore will have rather large errors. Our data therefore imply that, within the errors, the emission line
sources are associated with the clusters.

As can be seen in Figure 11, which shows our two dimensional spectra for GC6 and GC7 around 5000\AA, the
bright [\ion{O}{3}] emission is spatially contained within the clusters. In both clusters, there is some hint of
diffuse [\ion{O}{3}]$\lambda$5007 emission, but at a significantly higher velocity than the concentrated emission
in the clusters. (This may be the diffuse [\ion{O}{3}] emission observed at 985 km s$^{-1}$ with SAURON integral 
field unit spectroscopy by Sarzi \etal 2006.) We therefore conclude that the sources of the [\ion{O}{3}] emission are 
within GC6 and GC7.

\subsection{ELGC Morphology}

Interestingly, GC6 and GC7 are adjacent to one another on the sky, with a projected separation of 13.6\arcsec\
($\sim$850 pc at a distance of 13 Mpc). Their similar positions and radial velocities could suggest that these
two GCs are on similar orbits around the galaxy. Their positions can be seen in Figure 12, an \emph{HST}/WFPC2
image of NGC 7457.  The inset details in Figure 12 show GC6 and GC7 at WFPC2 resolution, along with normal
globular cluster GC8.

Some asymmetry is obvious for GC6, with the cluster appearing as an unresolved source with an extended tail in
both the F555W and F814W bands. It is possible that this GC is superimposed on a background object, with a more
distant galaxy appearing as the tail. However, with an upper limit to its size of R$_{eff} \lesssim$ 1 pc given
by our \emph{ishape} analysis, the concentrated knot of GC6 is much smaller than typical GCs, usually 3-4 pc
(Larsen 2004; Jord{\'a}n \etal 2005).  

On the other hand, GC7 has an unusually large effective radius, 12.3 pc. Curiously, Larsen \& Richtler (2006)
find similarly large sizes for the clusters in their sample which host emission-line objects, though these are
all young clusters with ages $< 1$ Gyr.

The luminosities of these two ELGCs are typical of globular clusters. GC7 has an absolute V magnitude of $-8.2$,
and GC6 has M$_V$ = $-8.1$ . This is less than $\sim$1 magnitude brighter than the turnover magnitude for the
Milky Way GC luminosity function (GCLF) (M$_V$ = -7.4), and is well within 1$\sigma$ (1.2 mag; Harris 2001) of the peak of a typical
log-normal GCLF.

\subsection{Emission Line Spectra}

The [\ion{O}{3}]$\lambda\lambda$4959,5007 emission lines dominate the emission-line spectra for both ELGCs.  
However, a few other emission lines become apparent in GC6 and GC7 after closer inspection. In order to isolate
the emission lines, the GC spectra have been normalized, and a MILES SSP model was subtracted from them. Because
of the issues described in Section 5, we can not be sure which SSP would be the best fit to GC6 and GC7.  
Therefore, throughout our analysis, we compared GC6 and GC7 with two MILES models: one was a model with solar
metallicity and an age of 2.0 Gyr, in order to mimic the stellar population in the nucleus of NGC 7457, while the
other was typical of a metal-rich GC, with an age of 12.6 Gyr and metallicity [M/H] = -0.38. The reader should
note that our choice of model does not affect our measurements of [\ion{O}{2}] and [\ion{O}{3}] much, but does
impact our results for H$\beta$. It is also important to note that the sensitivity of the spectrograph drops
significantly at blue wavelengths, so our measurements of [\ion{O}{2}] suffer from low signal-to-noise.

We require models of the underlying stellar population for two reasons. First, in order to account for any
absorption underlying the emission lines, we measure the MILES equivalent width and the GC equivalent width over
the same bandpass, and take the difference of the two to determine the total equivalent width of the emission
line. Secondly, we use the MILES models to constrain the overall shape of the continuum, so that when we compare
line strengths at 4861\AA/5007\AA\ with those at 3727\AA, we will not be affected by dust extinction or errors in
the flux calibration.

Table 6 lists the qualitative strengths of lines which are seen in emission after MILES SSP models are subtracted
from these two objects. GC7 is a high-excitation nebula, with no observable [\ion{O}{2}] emission, but very
strong [\ion{O}{3}] emission (L$_{5007}$ = $(1.2 \pm 0.3)\times10^{36}$ erg s$^{-1}$), and visible [\ion{Ne}{3}]
emission.  Unfortunately, we were not able to derive nebular parameters for this object, because the Balmer
emission and lines produced by other ionization states of oxygen were not strong enough to measure (see Figure
13).

GC6 is a much lower excitation nebula than GC7, with significant [\ion{O}{2}] emission and weaker [\ion{O}{3}]
emission (L$_{5007}$ = $(2.4 \pm 0.7)\times10^{35}$ erg s$^{-1}$). The [\ion{O}{2}] , H$\beta$, and [\ion{O}{3}]
emission lines can be seen for both clusters in Figure 13, with the solar metallicity, 2.0 Gyr MILES model
subtracted from them. The emission-line ratios for GC6 are listed in Table 7. Two values are listed for each
ratio: one utilizing the 2.0 Gyr MILES model, and one which uses the 12.6 Gyr old model. As we will discuss in
the next section, these line ratios for GC6 are difficult to explain in an old stellar population.

\subsection{The Nature of the Emission Line Sources}

Any thorough picture of the GC system of NGC 7457 must take these peculiar emission-line objects into account,
but we are forced to admit that there is no simple explanation for them. In extragalactic systems, a compact
source of strong [\ion{O}{3}] emission is often assumed to be a planetary nebula (PN). However, PNe have proven
elusive in GCs, and are still poorly understood. We must consider all the possible options for sources that might power
the observed emission lines.

In the Milky Way, only four PNe have been detected in 133 GCs surveyed (Jacoby \etal 1997). Meanwhile, in the large samples
of extragalactic GC spectra which now exist in the literature, only a small handful of ELGCs have been found.  
Little effort has been spent characterizing the sources of the observed emission lines. Three emission-line
objects were found in GCs in the giant elliptical galaxy NGC 5128 (Minniti \& Rejkuba 2002;  Rejkuba \etal 2003),
and were assumed to be PNe. Brodie \etal (2005) identified one ELGC in their spectroscopic sample of 22 GCs in
the elliptical galaxy NGC 4365, while in NGC 3379, Pierce \etal (2006) found that two of their 22
spectroscopically-observed GCs hosted emission line objects. Larsen \& Richtler (2006) performed a detailed study
of the [\ion{O}{3}]$\lambda\lambda$4959,5007 and [\ion{N}{2}]$\lambda\lambda$6548,6584 emission they found in
three young star clusters (aged between 30 Myr and 280 Myr) in the spiral galaxies NGC 5236 and NGC 3621.  
Finally, Zepf \etal (2007) recently found [\ion{O}{3}] emission in a NGC 4472 GC, and identified its source as a
black hole X-ray binary.

\subsubsection{Planetary Nebulae?}

It is difficult to constrain the ionization source creating the emission lines in GC7, as it lacks measurable 
[\ion{O}{2}]$\lambda$3727 and H$\beta$. However, the presence of [\ion{Ne}{3}] and strong
[\ion{O}{3}] are typical of PNe (Jacoby \& Ciardullo 1999). This source has an [\ion{O}{3}] absolute magnitude of
M$_{5007}$ = $-3.7$, as defined in Ciardullo \etal (2005). This is a typical value for a PN, 0.8 magnitudes
fainter than the PN luminosity function cut-off. We can get a rough lower limit for the mass of the central star
if we can derive lower limits for the luminosity and temperature of the central star. We constrain luminosity by
assuming that no more than 10\% of the central star's flux can emerge as the [\ion{O}{3}]$\lambda$5007 line (an
assumption supported by both theoretical and observational findings: Ciardullo \etal 2005). Therefore, the
central star of the PN in GC7 has a minimum luminosity of 3080 L$_\sun$. Additionally, we can place a lower limit
on the effective temperature of the central star if we can calculate E, the excitation parameter which is defined
by Dopita \etal (1992) as E~=~0.45~(~[\ion{O}{3}]$\lambda$5007~/~H$\beta$~). This formulation for E is only valid
if E $\le$ 5.0; to constrain greater values of E, a flux for the \ion{He}{2}$\lambda$4686 line must be measured.
Although we can not confidently measure H$\beta$, we can place upper limits on its flux, giving
[\ion{O}{3}]$\lambda$5007~/~H$\beta \gtrsim 15$, and clearly implying E $> 5.0$. Although we can not actually
measure E because the \ion{He}{2} line was not observable, we can use this lower limit on E to derive a lower
limit on the central star's temperature. Using equation 2.2 in Dopita \etal (1992), we find that log(T$_{eff}$)
$>$ 5.0.

As can be seen in Figure 14, which plots Vassiliadis \& Wood (1994) model tracks for post-AGB evolution, these
modest lower limits on the central star's luminosity and temperature imply a mass $> 0.58 M_{\odot}$,
corresponding (in these models) to a progenitor main sequence mass $>$1.1 M$_{\odot}$. These models are for
approximately solar metallicity; lower metallicity model tracks shift slightly upward, and the inferred lower
limit to main sequence mass would be $\sim$1 M$_{\odot}$ for [$Z$/H]=$-0.6$. In a solar metallicity population, a
turn-off mass of 1.1 M$_{\sun}$ corresponds to an age of 7 Gyr, while in a typical metal-poor GC, a 1.0
M$_{\sun}$ turns off at this age (Demarque \etal 2004); this implies that GC6 is $\le7$ Gyr old.

Intermediate ages are also supported by empirical studies; Jacoby \& Ciardullo (1999) derive nebular parameters
from high-quality spectra for PNe in M33, and find that all the PNe in their sample with M$_{5007}$ brighter than
$-3.0$ have central star masses $\geq$ 0.6 M$_{\sun}$. The white dwarfs of mass $0.61 M_{\odot}$ in the solar
metallicity open cluster NGC 7789 have progenitor masses of $2.0 M_{\odot}$ (Kalirai \etal~2007), so we would
expect a minimum progenitor mass of $\sim 1.5-2.0 M_{\odot}$ in our case. Therefore, both the theoretical and
empirical models indicate a progenitor main sequence mass of $> 1 M_{\odot}$ for GC7. If GC7 is an old GC, the
progenitor must be more massive than the main sequence turnoff stars---probably a blue straggler. If the
progenitor is instead a normal turnoff star, then the corresponding cluster age depends upon the exact mass, with
a maximum age of 7 Gyr and a more probable age of $\sim$2 Gyr. Such ages are consistent with the
luminosity-weighted age of the galaxy center.

On the other hand, the emission line spectrum of GC6 is of such low excitation that it would constitute a rather
unusual PN. The near-equal strengths of [\ion{O}{2}] and [\ion{O}{3}] imply a cool central star (T $\approx$
36,500 K), implying that much less than 10\% of the central star's flux is observed as [\ion{O}{3}]. Given the
significant [\ion{O}{3}]$\lambda$5007 luminosity measured, one finds that the luminosity of the central star
would have to be at least 3$\times$10$^{4}$ L$_\sun$. These parameters imply a central star with a core mass of
$>$0.9~M$_{\sun}$ (Vassiliadis \& Wood 1994). As is apparent from Figure 14, the progenitor of such a central
star would have been very massive on the Main Sequence-- greater than 5 M$_\sun$. Such a massive star is
difficult to explain in any but the youngest stellar populations. A blue straggler of such mass is
extraordinarily unlikely. Therefore, the emission line source in GC6 is most probably not a PN.

\subsubsection{Supernova Remnants?}

CJFN 470 is an emission line
source in M31 which closely resembles that in GC6, and was classified as a PN by Jacoby \& Ciardullo (1999). It
is the only source in their sample with approximately equal strengths of [\ion{O}{2}]$\lambda$3727 and
[\ion{O}{3}]$\lambda$5007. A closer look at this source reveals [\ion{S}{2}] lines which are nearly as strong as
H$\alpha$---a strong indicator that this object is a SNR.

Bright lines from lower ionization states are expected if the nebula is shock-heated, and not
photoionized. Supernova remnants (SNRs) are often selected by their relatively bright [\ion{S}{2}] emission (as
compared with H$\alpha$), and display high levels of [\ion{O}{1}] and [\ion{O}{2}] emission as well (Fesen \etal
1985). It is probable that GC6 hosts a SNR, although it is difficult to be certain without spectral coverage of
the [\ion{S}{2}]$\lambda\lambda$6717,6731 and [\ion{O}{1}]$\lambda\lambda$6300,6363 lines.

A detection of an SNR in a GC by its [\ion{O}{3}] lines is not unprecedented. Minniti \& Rejkuba (2002) asserted
that they had found the first PN in a GC in an elliptical galaxy. However, with the higher-resolution
spectroscopy of Peng \etal (2004), the [\ion{O}{3}] emission lines were observed to be double-peaked, with the
peaks separated by $\sim$300 km s$^{-1}$. The authors note that such kinematics are implausible for a PN, which
usually have maximum expansion velocities of a few times 10 km s$^{-1}$.

There is no evidence for similarly high expansion velocities in our emission line objects. In both cases, the
[\ion{O}{3}] lines are unresolved, placing an upper limit on their velocity widths of 165 km s$^{-1}$. However,
this upper limit does not exclude the possibility of GC6 hosting an SNR. The oxygen lines in the spectrum 
imply that the SNR is in an evolved, radiative stage with emission arising from swept-up shocked ISM. Typical  
shock velocities for this stage are 50-200 km s$^{-1}$ (Weiler \& Sramek 1988).

Maoz \etal (2005) point out that an SNR in a very low-density medium will only glow as a non-radiative Balmer-dominated 
SNR, and once it has expanded through all of its circumstellar material, it will fade quickly. Therefore, the presence of radiative 
SNRs in GCs around early-type galaxies can place constraints on the ISM density in these clusters. Future observations with 
better velocity resolution and extended wavelength coverage will allow us to model the gas density in GCs.

This SNR probably does not explain the curious morphology of GC6 as it is quite evolved. The optical continuum emission of 
a SN Type Ia fades quickly, falling to the R-band luminosity of GC6 2-3 years after the explosion, and to 10\% the luminosity of 
GC6 less that one year after that (Woosley \etal 2007, Lair \etal 2006). 

Although an unrelated bright point source might contribute to the compact profile of the cluster and the illusion of a lower-surface 
brightness tail, we can exclude transient phenomena as the cause of GC6's morphology. The flux of GC6 does not appear to be 
time-variable, as the spectroscopic continuum flux of GC6 in our 2003 observations is approximately the same as or other GCs 
which had similar V magnitudes as GC6 in the Cycle 4 HST images.

\subsubsection{Binary Stars?}

Several kinds of binary star systems can produce bright [\ion{O}{3}] emission in intermediate-to-old stellar
populations. One such breed of binary system are symbiotic stars (SSs), which feature a hot component (usually a
white dwarf), and a cool component (an evolved giant). The dense wind from the giant is irradiated with UV
emission from the white dwarf, producing emission lines which can closely mimic those in PNe. Within the Local
Group, SSs can be differentiated from PNe by the underlying red continuum from the giant. However, at distances
beyond the Local Group, the continuum from this single star becomes too faint to detect (Magrini \etal 2003).
Additionally, emission-line sources in GCs are surrounded by stars, which overwhelm any stellar giant companion
to a white dwarf. We must therefore use emission line ratios to distinguish PNe from SSs.

Guti{\'e}rrez-Moreno \etal (1995) considered such spectral diagnostics, and determined that SSs do not show
[\ion{O}{3}]$\lambda$5007/H$\beta\ \gtrsim$ 10. Objects with [\ion{O}{3}]/H$\beta\ <$ 10 may be either SSs or
PNe.  As discussed in the last section, for GC7 we know that [\ion{O}{3}]/H$\beta >$ 15. Therefore, GC7 probably
hosts a PN.

We can not use this diagnostic for GC6, because its [\ion{O}{3}]/H$\beta$ = 2.3-3.9, well within the regime
inhabited by both PNe and SSs. However, diagnostic diagrams developed in Schwarz (1988) and Guti{\'e}rrez-Moreno
(1988) make use of [\ion{O}{2}]$\lambda$3727, [\ion{O}{3}]$\lambda$5007, and H$\beta$, and are useful for
constraining the nature of GC6. These diagrams employ the $\Delta$E parameter defined in Baldwin \etal (1981),
where: $$\Delta E = \textrm{log( [\ion{O}{3}] / }H\beta\ ) + \textrm{log( ( [\ion{O}{2}] / [\ion{O}{3}] )} +
0.32\ ) - 0.44.$$ We measure values of $\Delta$E = 0.1-0.3 for GC6 (depending on which SSP model is subtracted).
Symbiotic Stars with similar [\ion{O}{2}]/[{\ion{O}{3}] as GC6 have $\Delta$E $<$ -0.15, so it is clear that GC6
does not contain a SS. Curiously, GC6 sits comfortably in the \ion{H}{2} region zone of the Schwarz (1988)
diagnostic plot. However, we dismiss this option because our measured absorption-line spectrum clearly shows that
GC6 does not contain young ($<$ 10 Myr) stars.

Finally, a binary system containing a compact object can generate optical emission lines (Zepf \etal 2007). We
examined archival \emph{Chandra X-ray Observatory} data for NGC 7457, and concluded that there are no X-ray
sources coincident with GC6 or GC7. Additionally, the low velocity dispersions of the emission lines in these
clusters would be unusual for a binary system containing a neutron star or black hole (for contrast, see the
broad velocity structure of the [\ion{O}{3}] lines in Zepf \etal~2007).

We can therefore exclude the possibility of GC6's emission coming from a PN, an SS, an X-ray binary, or an
\ion{H}{2} region; the emission lines are probably coming from an SNR.

\section{Discussion and Conclusions}

Together, these analyses strongly suggest a significant population of intermediate-age GCs in NGC 7457. Although
no one piece of evidence is conclusive, there is a large body of data hinting towards peculiar stellar
populations in this GC system. In review, the evidence suggesting intermediate-age GCs includes:

\begin{enumerate}

\item V-I color-magnitude diagram. Even with \emph{HST} photometry, the distribution of GC colors appears broad,
and does not have two subpopulations at the ``usual" locations. A simple explanation for this is 
an intermediate-age stellar population filling in the gap 
between the two normal old GC subpopulations.

\item Color-color analysis. The location of several objects in the $F330W-V$ vs.~$V-I$ plot are consistent with 
intermediate ages, but do not require them.

\item High metallicities of GC6 and GC7. The metallicities of these two clusters are significantly higher than 
any of the other GCs, and in many of our measurements they appear more similar to the galactic nucleus than to 
typical metal-rich GCs.

\item Strong higher-order Balmer line absorption for GC6 and GC7. H$\delta$F should be less affected by 
emission-line fill-in than H$\beta$, and GC6 and GC7 are found near the 2.5 Gyr-old galactic nucleus in an 
H$\delta$F-$<$Fe$>$ line diagnostic plot.

\item Significant evidence for a $>$1 M$_{\sun}$ progenitor for the PN in GC7. Although such a progenitor could have 
been a blue straggler in an old stellar population, it would be simpler to explain it as a normal main sequence star in a 
$\sim 2-3$ Gyr old stellar population.

\item A probable SNR in GC6. Given the intermediate-to-old nature of the GC stellar population, the SN was almost 
certainly Type Ia. Various models featured in Matteucci \& Recchi (2001) predict that SNe Ia rates are anywhere 
between 7 and 158 times higher in a 2.5 Gyr old stellar population as compared to a 12 Gyr old stellar population.

\end{enumerate}

All together, this evidence makes a strong case for the existence of at least some intermediate-age GCs, which 
trace the star-formation event that dominates the stellar population in the center of NGC 7457. The total number 
of such GCs is uncertain. Two out of 13 spectroscopic GCs may be of intermediate age, while photometry suggests 
that a third or more of the GCs in the WFPC2 frame could be of intermediate-age. Much larger spectroscopic samples 
would be required to determine the fraction conclusively.

It is important to remember that the two near-universal GC subpopulations of old metal-rich and old metal-poor 
stars are present in NGC 7457. There are significant numbers of GCs at the appropriate colors predicted by the GC 
color-galaxy luminosity relation. Meanwhile, eleven out of 13 spectroscopic GCs appear old, with metallicities 
ranging from [Fe/H] = $-0.8$ to $-1.8$.

The kinematics of the GCs show rotation around the photometric minor axis of the galaxy. The orientation of the 
rotation is generally consistent with that of the galaxy itself, as determined by integral-field spectroscopy.

What general conclusions may be drawn about the history of this unusual S0? NGC 7457 has been seen as a classic 
example of a pseudobulge (that is, a disky bulge built through secular processes) due to its very low velocity 
dispersion. The existence of a significant number of old, metal-rich GCs---usually associated with merger-built 
``classical'' bulges---indicates that the bulge of NGC 7457 is unlikely to have been built solely through secular 
means at relatively recent times. A number of GCs appear to have been added in a recent star-forming event that 
dominates the stellar population of the most central regions; therefore, the event probably involved a substantial 
amount of gas. 

Even the younger GCs need not be associated with a pseudobulge. The two ELGCs with presumptive
intermediate-ages are located at projected radii of 32\arcsec\ and 44\arcsec\ (2.0
and 2.8 kpc). The bulge itself dominates the light profile only within 9\arcsec\ (J.~Kormendy 2008, private
communication), and it is not clear that purely secular star formation would lead to GC formation at larger
radii.

A larger kinematic and stellar population study of the metal-rich GCs in NGC 7457 would help illuminate the 
relative importance of rotation and pressure support among the old and intermediate-age GCs, and thus whether a 
secular or merger origin is more likely.

We have also shown that it is critical to acknowledge and understand the presence of emission lines in spectroscopic 
GC samples, as emission can strongly affect some lines (such as Balmer lines) through fill-in and thus change the 
derived cluster ages. Our analysis has also demonstrated the utility of emission line objects as an alternate route 
to studying the stellar populations of extragalactic GCs. While we did not have the necessary emission lines in this 
study to calculate chemical abundances, future studies with higher S/N spectra and wider wavelength coverage could 
estimate abundances of light elements like O, N, and S that are difficult to derive from absorption-line spectra.

\section{Acknowledgments}
  
We are very grateful to Jay Gallagher, George Jacoby, John Kormendy, and Nick Konidaris for useful discussions and we 
thank our referee, Jimmy Irwin, for his helpful suggestions. We are indebted to Rob Proctor for insight into his multi-index fitter. 
We also thank Javier Cenarro, Glenda Denicol{\'o}, Mike Beasley, and Duncan Forbes for their help with the data reduction. 
We acknowledge support from the National Science Foundation through grants AST 0206139 and AST 0507729. Support for 
this work was provided by NASA through Hubble Fellowship grant HF-01214.01-A awarded by the Space Telescope Science Institute, which is
operated by the Association of Universities for Research in Astronomy, Inc., for NASA, under contract NAS 5-26555.

This research has made use of the NASA/IPAC Extragalactic Database (NED) which is operated by the Jet Propulsion 
Laboratory, California Institute of Technology, under contract with the National Aeronautics and Space 
Administration. The SSC Spitzer Tools also proved useful to this research. Finally, we acknowledge the usage of 
the HyperLeda database (http://leda.univ-lyon1.fr).

\newpage
\includegraphics[width=14cm]{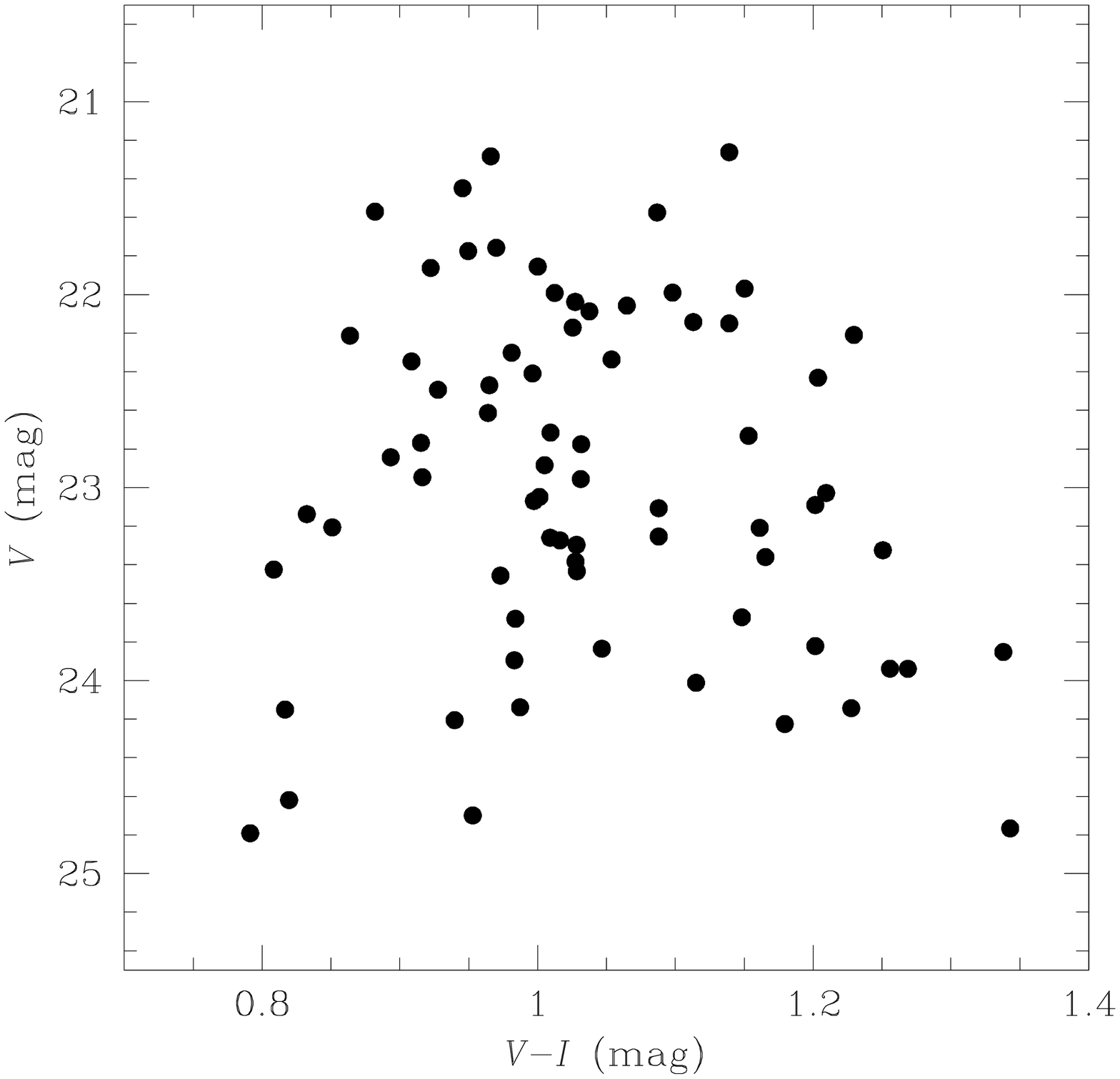}
\figcaption[cmd]{\label{fig:cmd} Color-magnitude diagram of GC candidates in NGC 7457 from $VI$ HST/WFPC2 imaging.}

\newpage
\includegraphics[width=14cm]{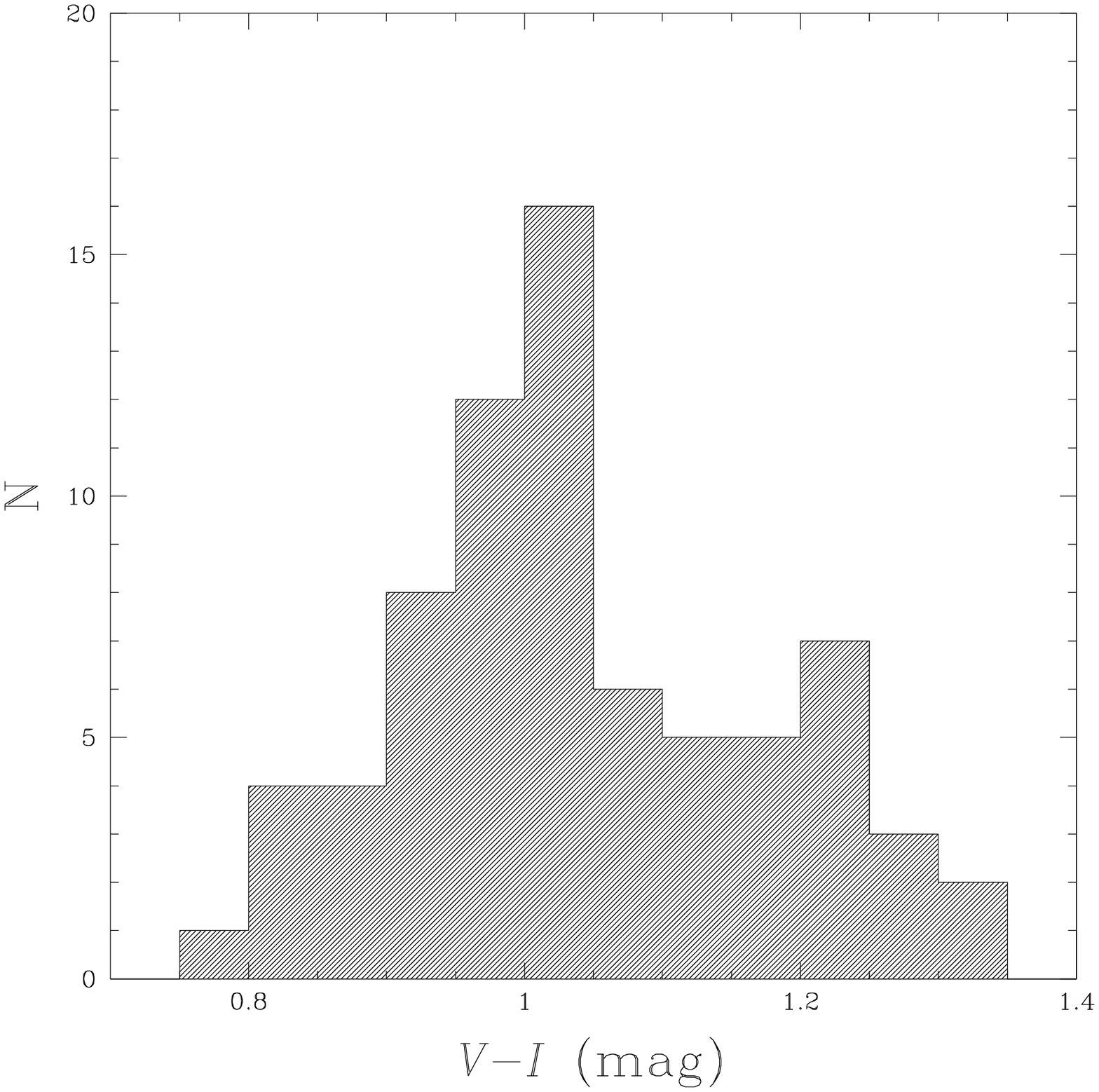}
\figcaption[hist]{\label{fig:hist} $V-I$ histogram of GC candidates. There is evidence for multiple subpopulations of
GCs, and an enhancement of objects at $V-I \sim 1.02$ that is unusual for an early-type galaxy.}

\newpage
\includegraphics[width=14cm]{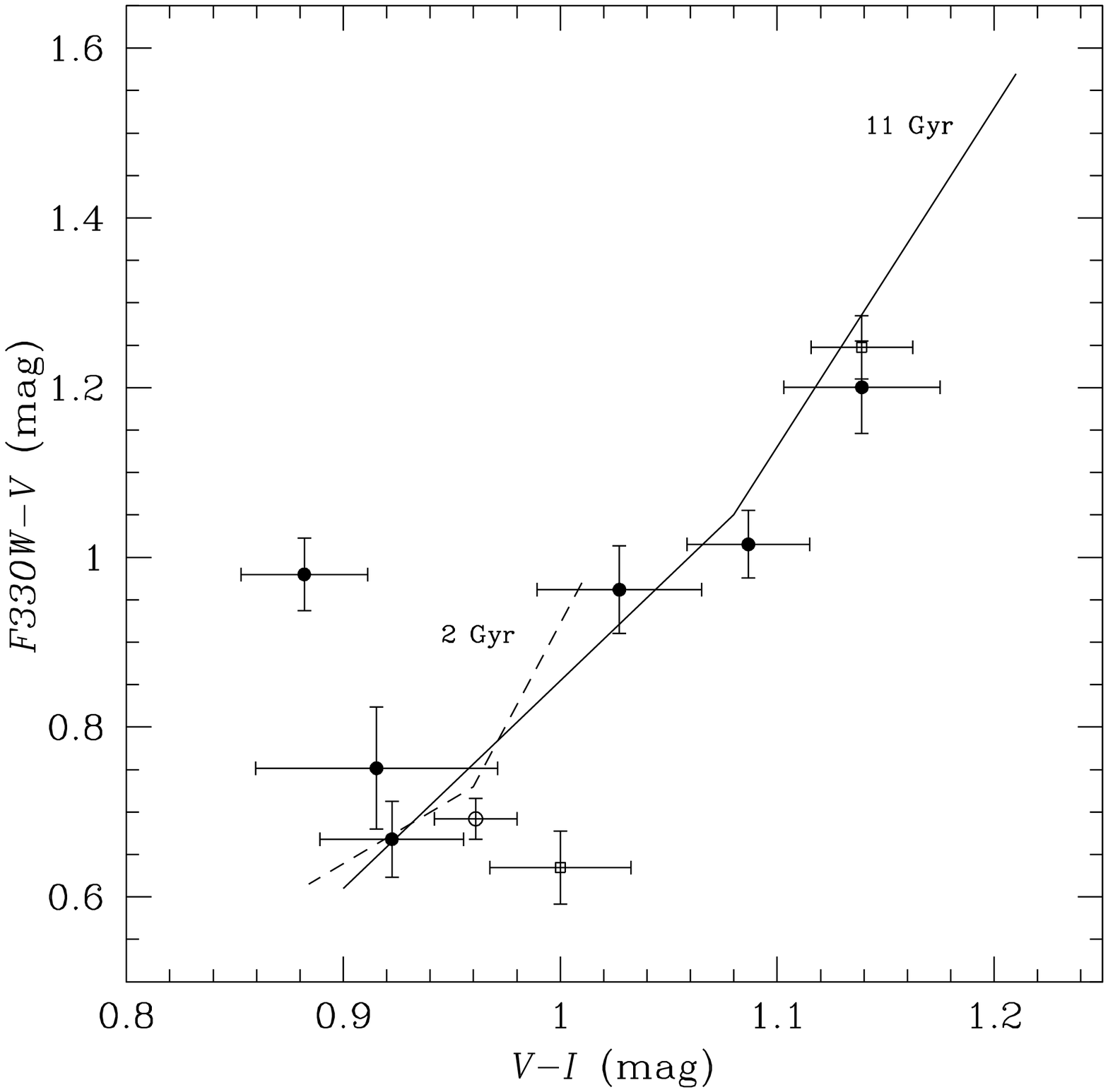}
\figcaption[uv2]{\label{fig:uv2} $F330W-V$ vs.~$V-I$ color-color diagram of GC candidates. The open boxes are GCs with spectroscopic 
confirmation, the open circle is the large and luminous object L1; the filled circles are the remainder of the candidates.
The dotted and solid lines are 2 and 11 Gyr isochrones from Girardi \etal~(2002) with metallicities from [$Z$/H] = $-1.7$ to 0.
A number of objects are too red to be of intermediate age, while several are consistent with either isochrone.}

\newpage
\includegraphics[width=14cm]{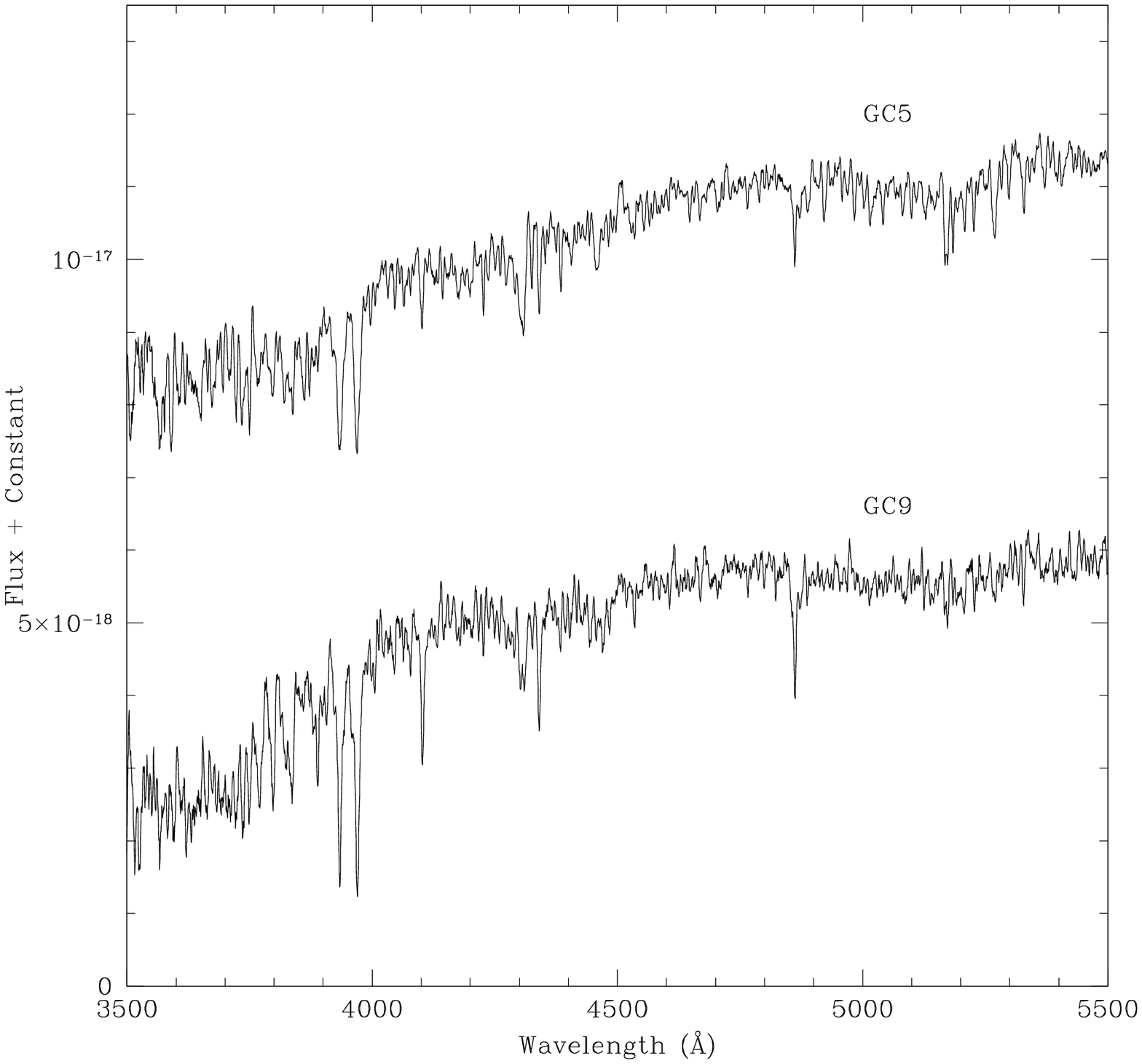}
\figcaption[plotspec]{\label{fig:plotspec} Spectra of metal-poor globular cluster GC9, and metal-rich globular cluster GC5. Notice the differences in strength of the Balmer line absorption (particularly H$\beta$ at 4861\AA\ and H$\delta$ at 4102\AA) and the Mg band at 5170\AA.} 

\newpage
\includegraphics[width=14cm]{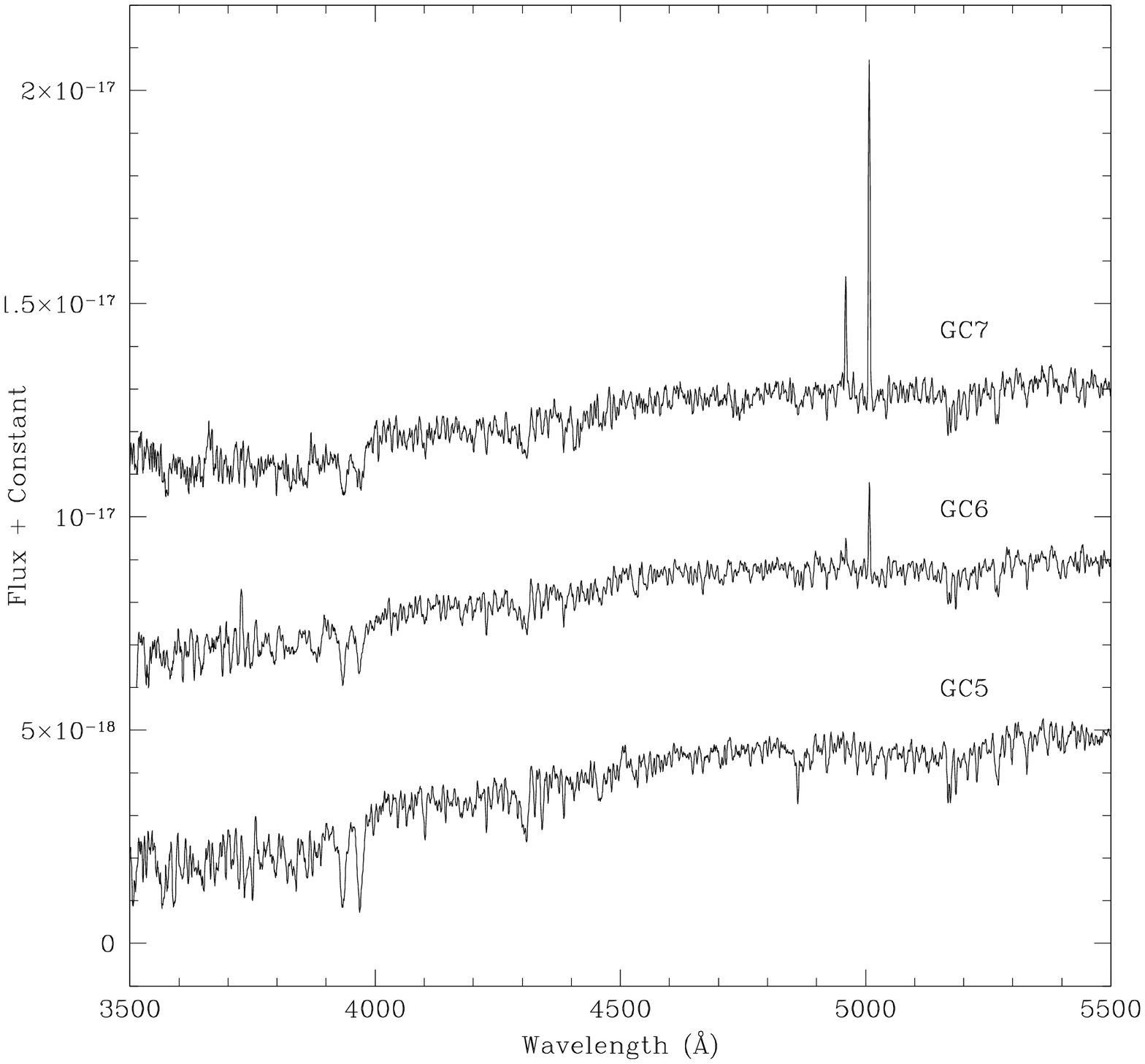} 
\figcaption[plotspec_pn]{\label{fig:plotspec_pn} Spectra of GC6 and GC7, the two globular clusters with bright [\ion{O}{3}]$\lambda \lambda$4959,5007 emission lines. Normal metal-rich globular cluster GC5 is included for comparison. GC6 features [\ion{O}{2}]$\lambda$3727 emission, and both GC6 and GC7 lack Balmer absorption due to fill-in by emission lines.} 

\newpage
\includegraphics[width=14cm]{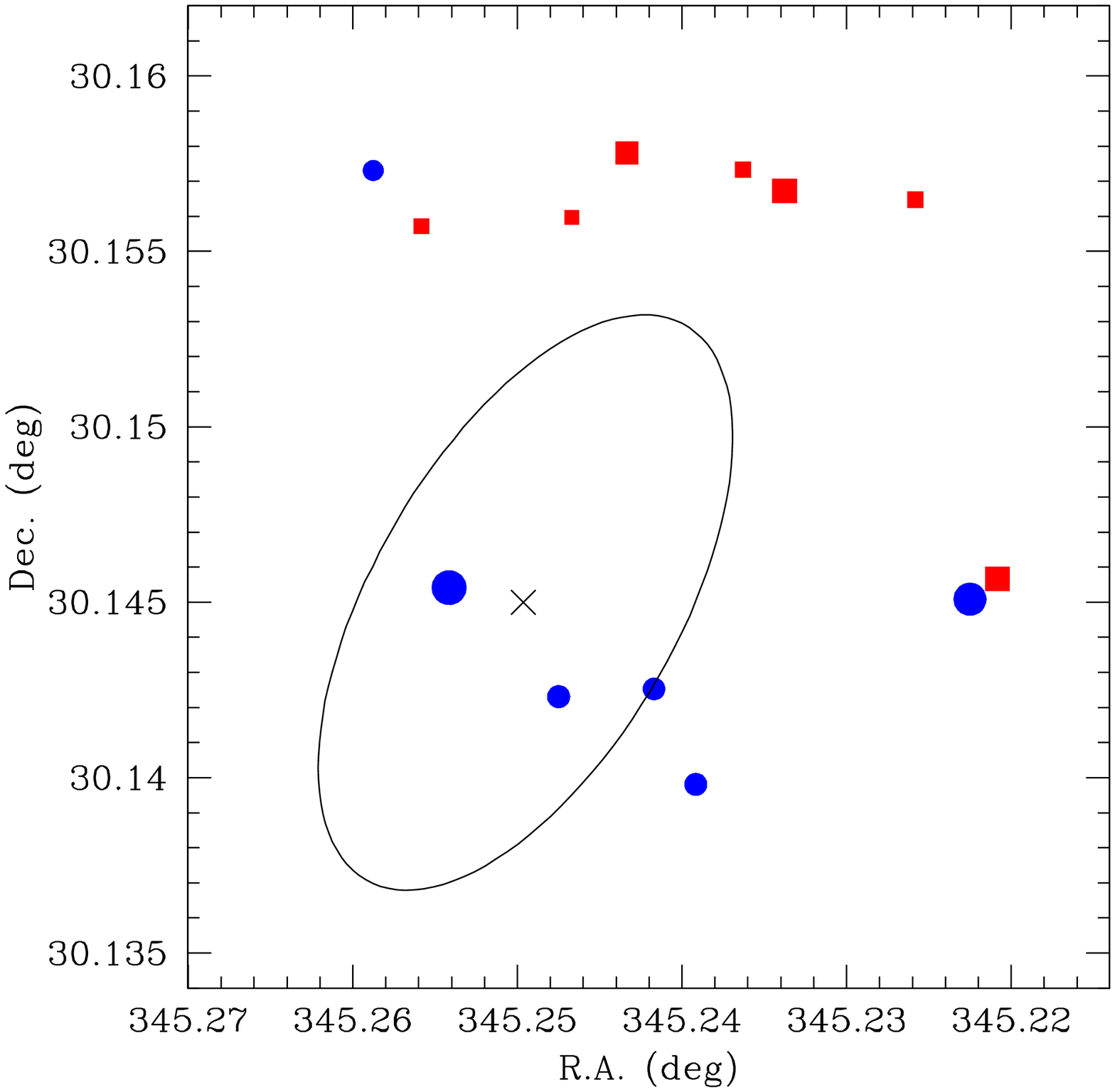} 
\figcaption[rv3]{\label{fig:rv3} Kinematics of GCs in NGC 7457. Blue circles are GCs with radial velocities less than the median for the system; red squares have velocities larger than the median. The size of the point is proportional to the deviation from the median velocity. The center of the galaxy is marked with a cross, and a representative photometric isophote is plotted. Despite the unusual spatial distribution of the GCs, the general impression is that the GC system is rotating around an axis similar to that of the photometric minor axis of the galaxy.}

\newpage
\includegraphics[width=14cm]{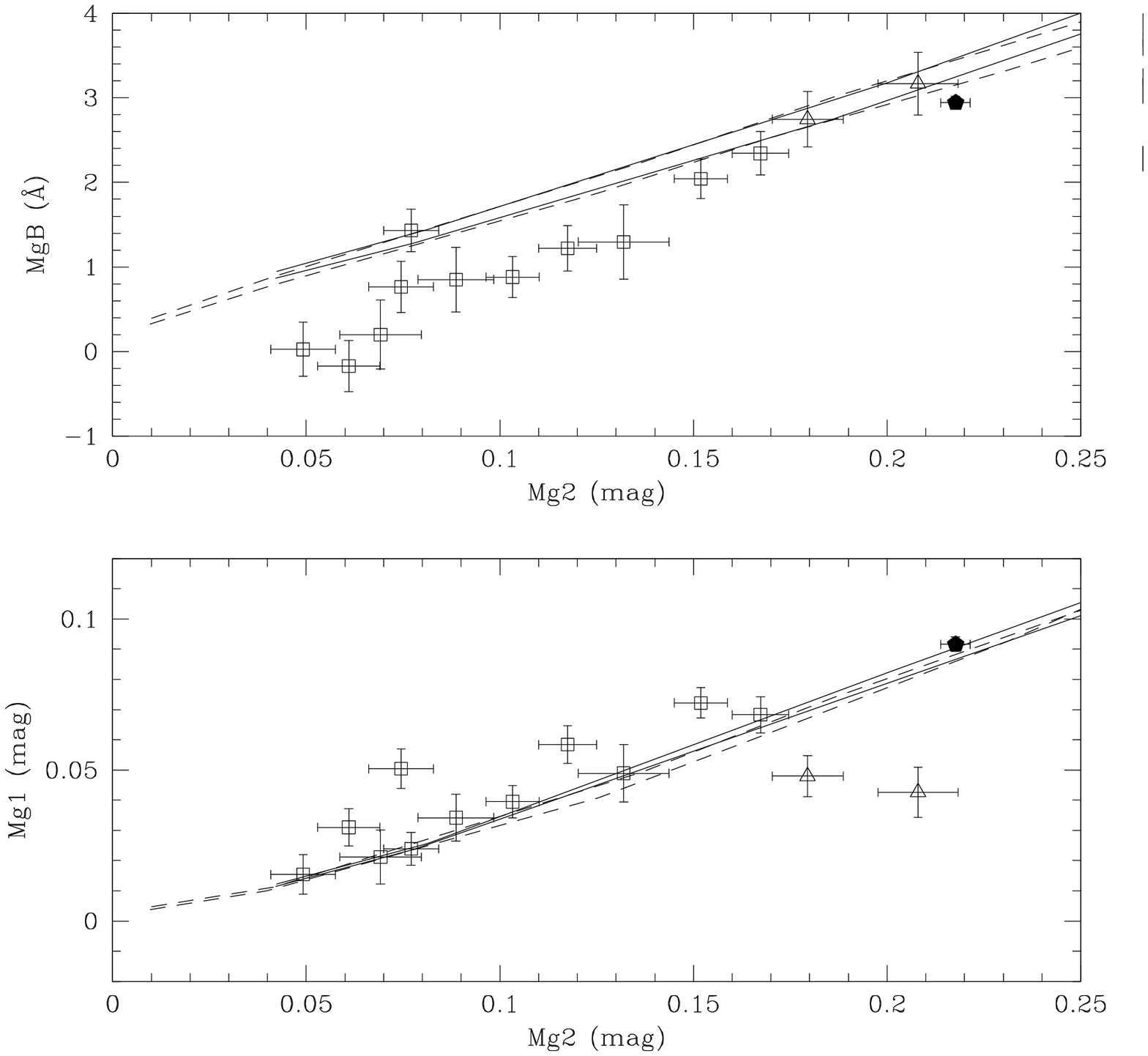}
\figcaption[mg_comp_ne]{\label{fig:mg_comp_ne} Figures comparing the Mg2 index with Mg$b$ and Mg1. Thomas \etal 2003 models are plotted as lines. Solid lines represent a sequence of metallicity for a 14 Gyr old stellar population. The dashed lines are models for a 2 Gyr old SSP. For each age, the two individual lines represent two different $\alpha$-element ratios ([$\alpha$/Fe] = 0.0 and 0.3). Open squares are our 11 normal GCs, while open triangles represent the two GCs with emission. The filled pentagon represents the center of NGC 7457 itself.}

\newpage
\includegraphics[width=14cm]{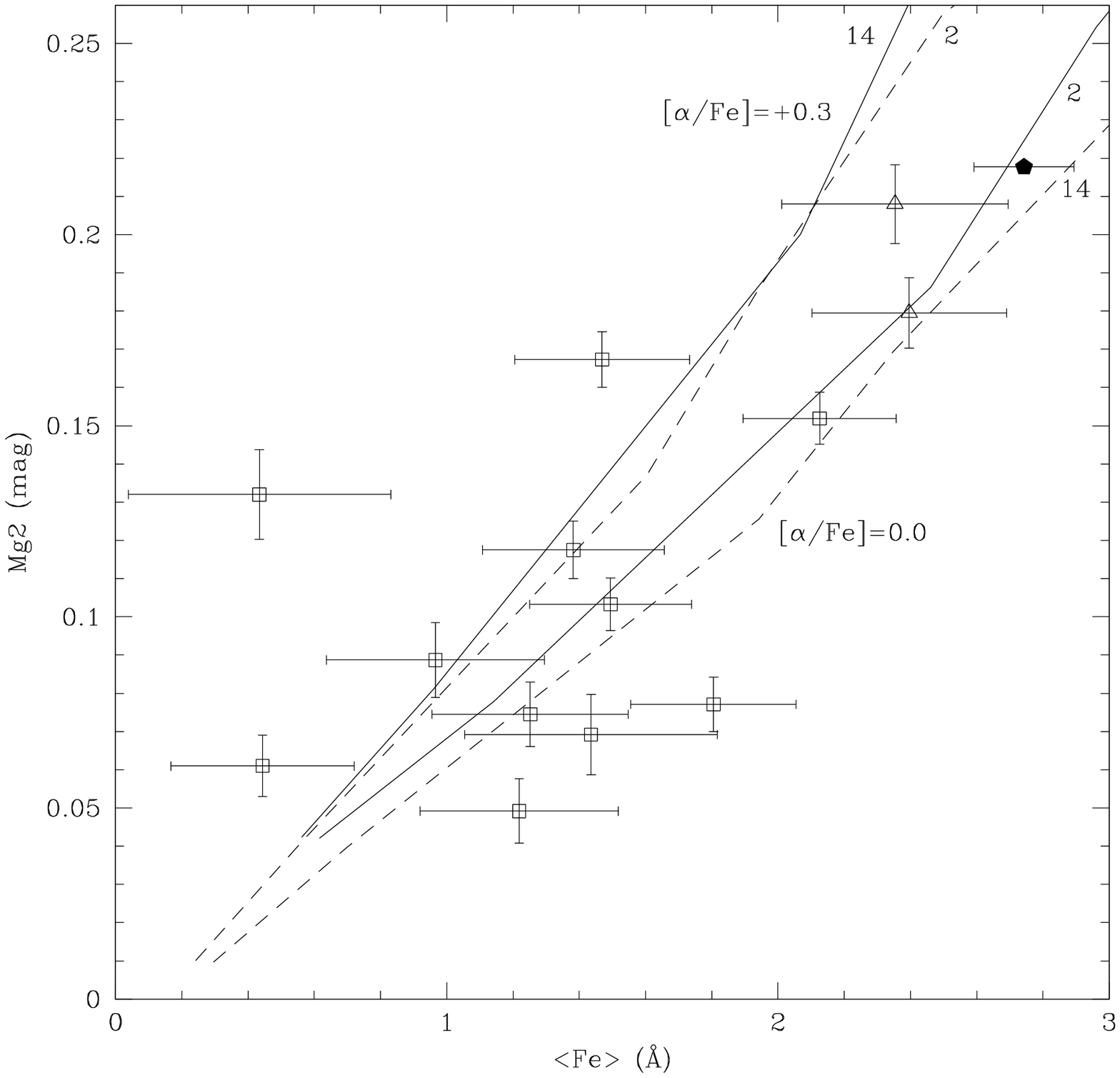}
\figcaption[mg2_fe]{\label{fig:mg2_fe} Plot of Mg2 vs. the $<$Fe$>$ index. The symbols and model lines are as in Figure 7. Model lines are labeled with ages and [$\alpha$/Fe].}

\newpage
\includegraphics[width=16cm]{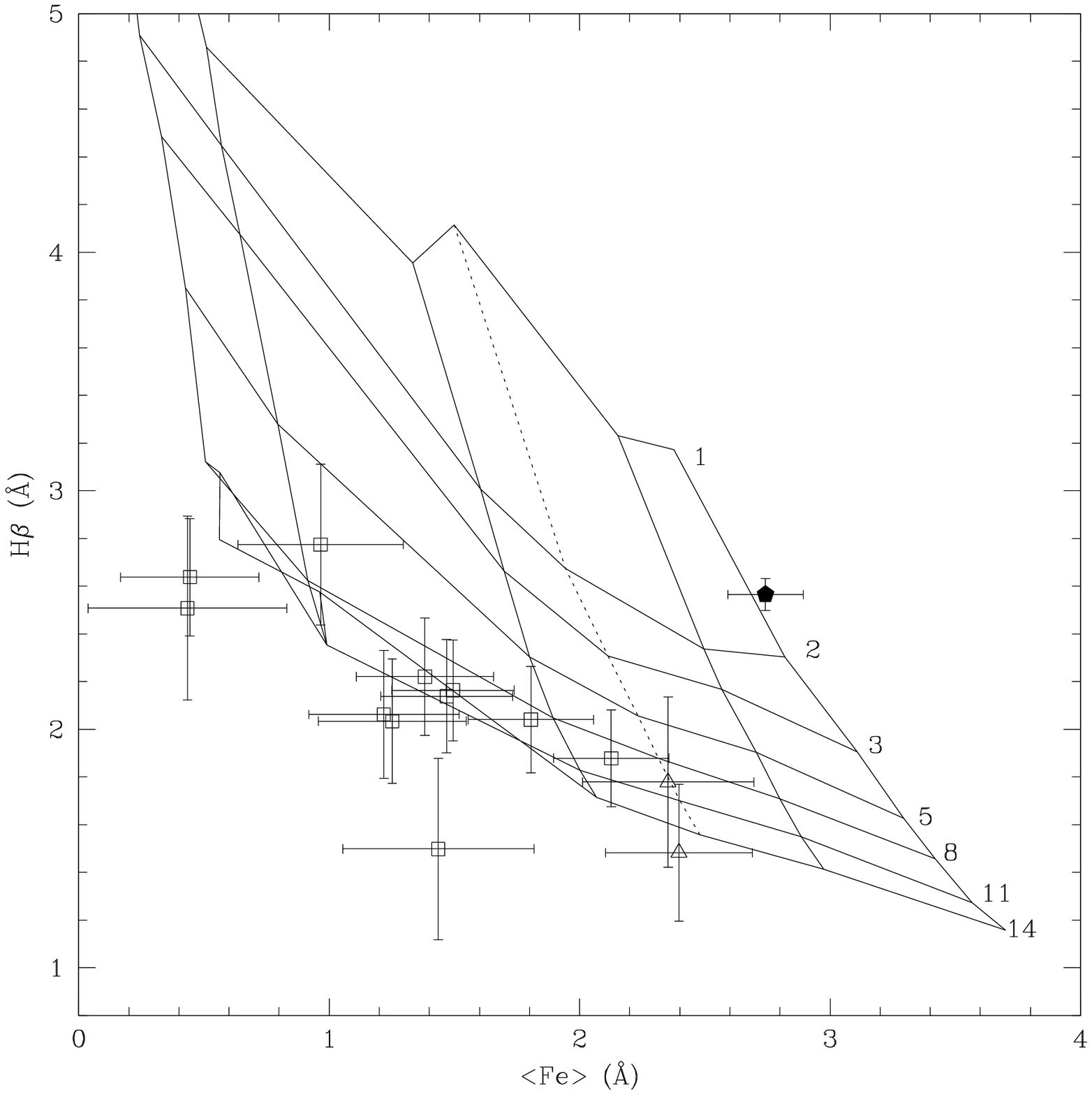}
\figcaption[hb_fe_ne]{\label{fig:hb_fe_ne} Plot of H$\beta$ vs. $<$Fe$>$ index, with a grid of model isochrone and isometallicity lines overlain (from Thomas \etal 2003, 2004). From left to right, the isometallicity lines represent [$Z$/H] = $-$2.25, $-$1.35, $-$0.33, 0.00, 0.35, 0.67. The solar isometallicity line can be seen as the dotted line. All metallicities assume [$\alpha$/Fe] = 0.3. Ages are indicated at the right of each isochrone, in Gyr. Open squares are our 11 normal globular clusters, while open triangles represent the two GCs with emission. The filled pentagon represents the center of NGC 7457 itself.}

\newpage
\includegraphics[width=16cm]{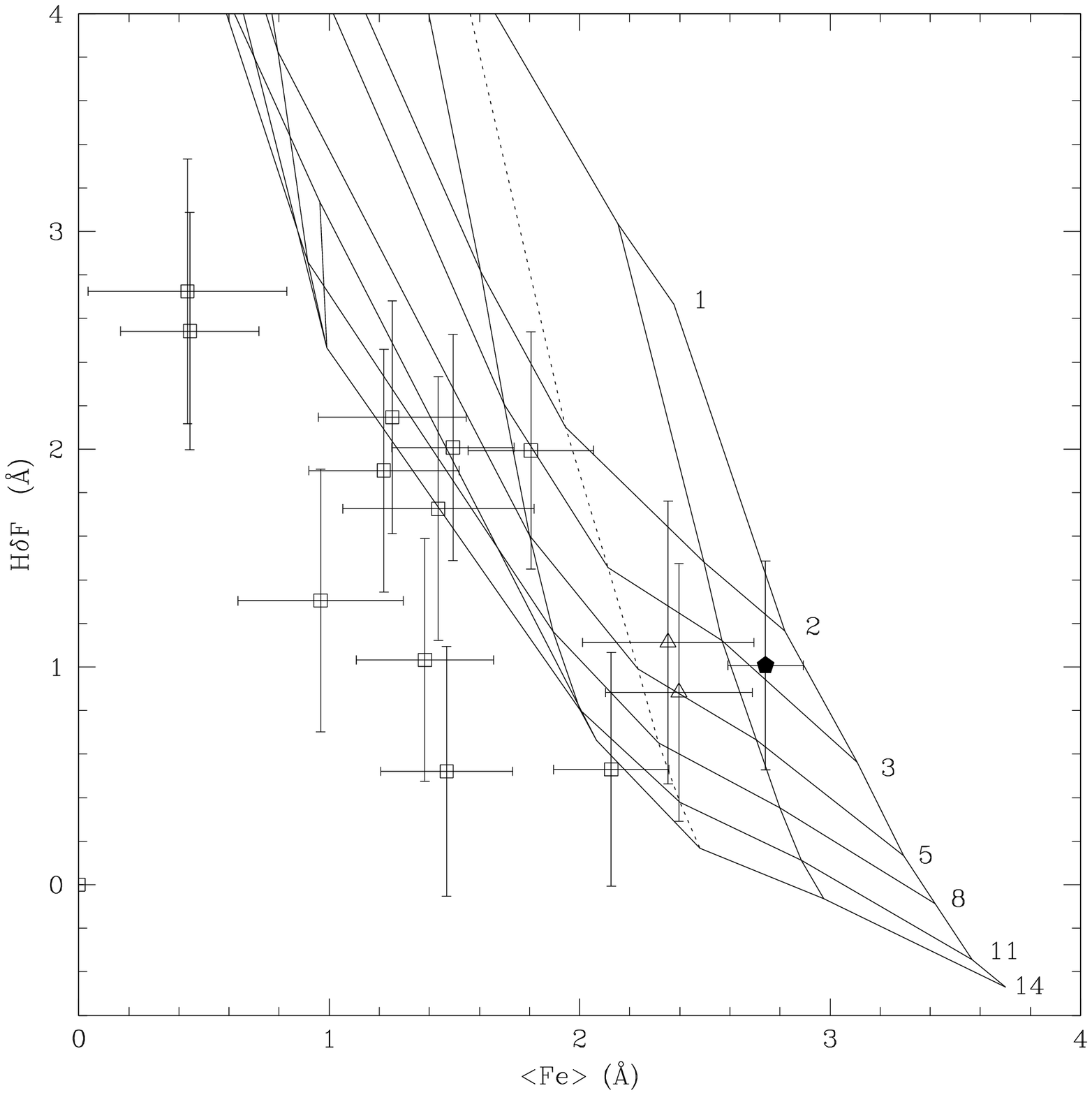}
\figcaption[hb_fe_ne]{\label{fig:hb_fe_ne} Plot of H$\delta$F vs. $<$Fe$>$ index. Symbols and grids as in Figure 9.}

\newpage
\includegraphics[width=15cm]{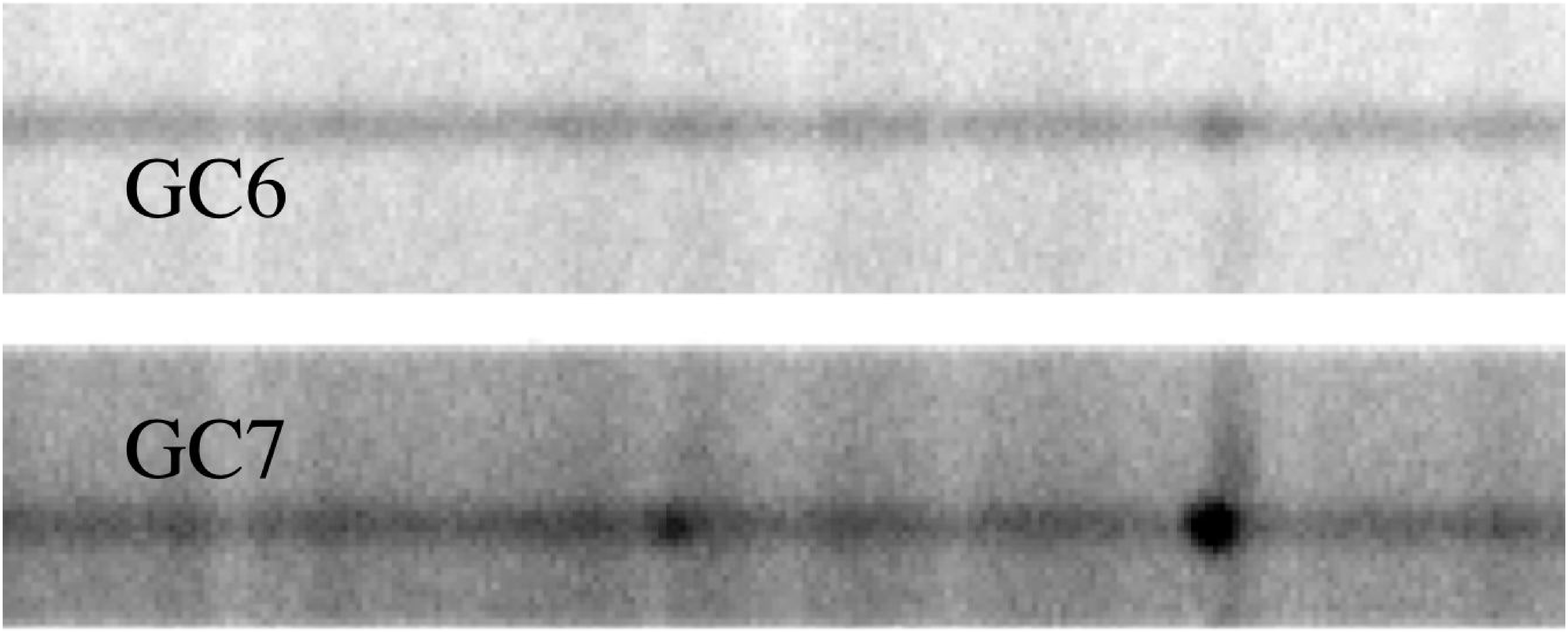}
\figcaption[o3_2d]{\label{fig:o3_2d} Two-dimensional spectra for GC6 and GC7, showing the [\ion{O}{3}]$\lambda\lambda$4959,5007 emission lines. Shorter wavelengths are to the left. The [\ion{O}{3}] lines in GC6 are significantly fainter than those in GC7, so the [\ion{O}{3}]$\lambda$4959 line is not visible here for this cluster. Notice the possible diffuse emission slightly redward of the compact [\ion{O}{3}]$\lambda$5007 emission.}
\vspace{1cm}
\includegraphics[width=14cm]{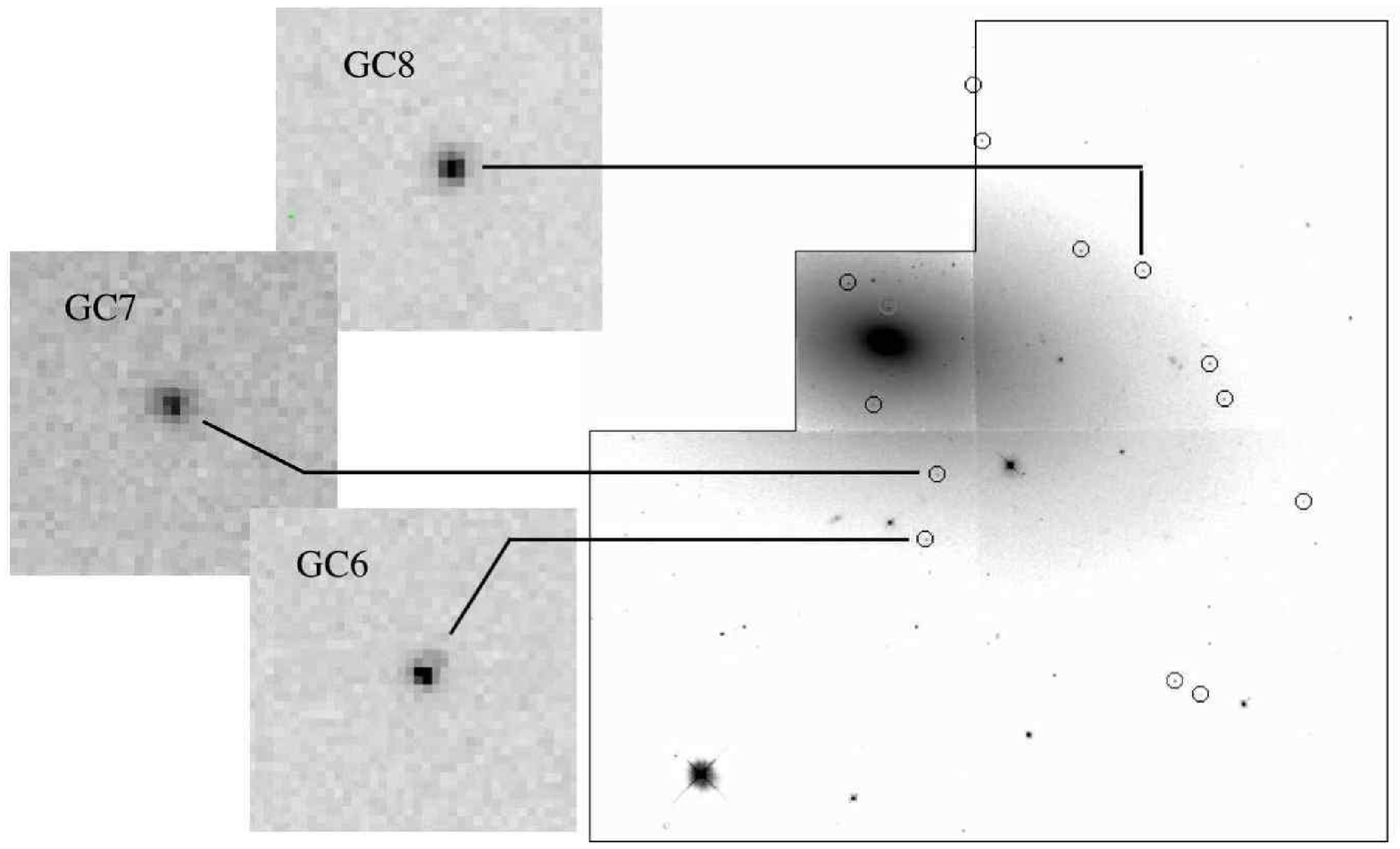}
\figcaption[pn2]{\label{fig:pn2} WFPC2 F814W image of NGC 7457, with our 13 spectroscopic GC targets marked by black circles. The two GCs which exhibit emission lines are shown in detail on the left, alongside GC8, a normal GC.}

\newpage
\includegraphics[angle=90,width=15cm]{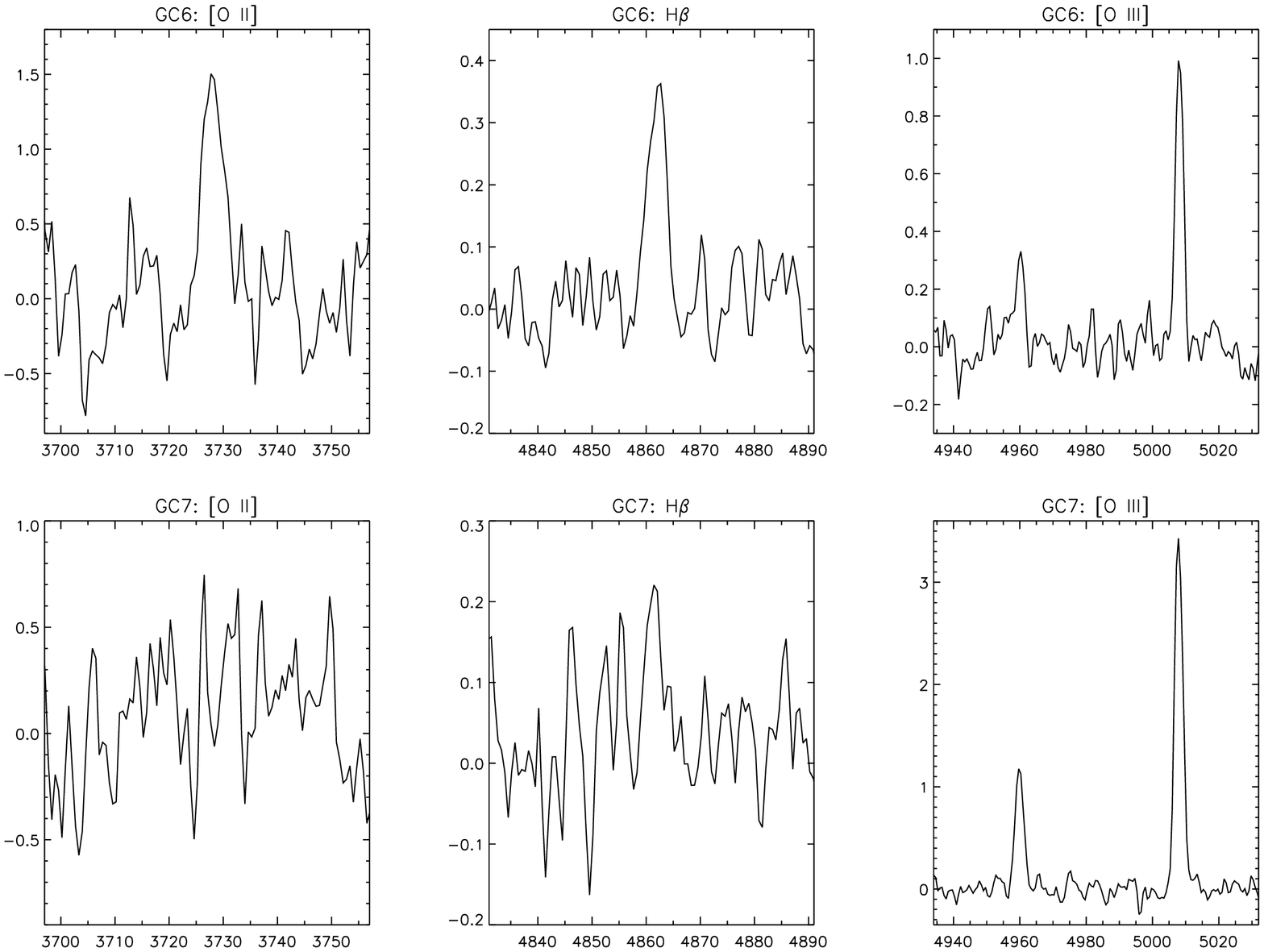}
\figcaption[emlines]{\label{fig:emlines} Detail of emission line spectra for GC6 and GC7. These are segments of the original spectrum, 
which have been normalized and had a 2.0 Gyr old, solar metallicity MILES SSP model subtracted from them. Clearly, H$\beta$ and 
[\ion{O}{2}] were too weak to be reliably measured in GC7.}

\newpage
\includegraphics[width=15cm]{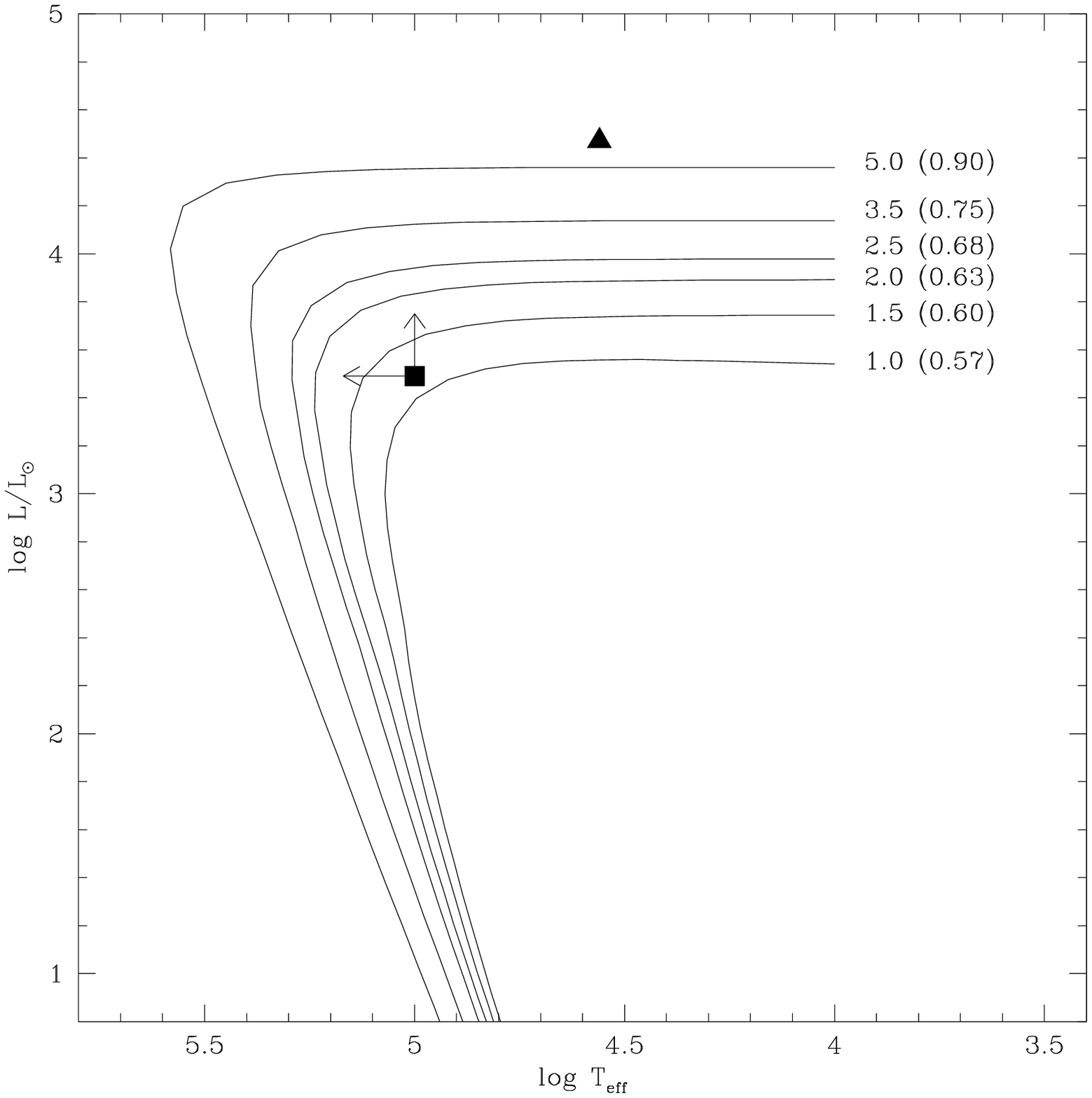}
\figcaption[pn_tracks]{\label{fig:pn_tracks} Model tracks for post-AGB stellar evolution from Vassiliadis \& Wood 1994, with GC6 overplotted as a triangle and GC7 represented by a square. Note that the plotted luminosity and temperature are both lower limits for GC7. The models are for metallicity Z=0.016, and each track is labeled with the mass of the progenitor main-sequence star, and in parentheses, the core mass at 10$^{4}$ K.}

\newpage
\begin{deluxetable}{lcc}
\tablewidth{0 pt}
\tablecaption{ \label{tab:gal}
  Global Parameters for  NGC 7457}
\tablehead{Parameter & Value & Source}

\startdata

Type  & SA(rs)0-? & NED \\
R.A. (2000) (h:m:s) & 23:00:59.9 & NED \\
Dec. (2000) (d:m:s) & +30:08:42 & NED \\
Radial Vel. (km s$^{-1}$) & 812 $\pm$ 6& NED \\
m - M & 30.55 $\pm$ 0.21 & Cappellari \etal 2006 \\
M$_{V}$ & -19.35 $\pm$ 0.24 & de Vaucouleurs \etal 1991\\
(B - V)$_{0}$ & 0.81 & HyperLeda\tablenotemark{a}  \\
A$_{V}$ & 0.17 & Schlegel \etal 1998 \\
Inclination (degrees) & 76 & HyperLeda\tablenotemark{a}  \\
Dust? & None & Peletier \etal 1999 \\
H$_{2}$ Mass (M$_{\sun}$) & (3.3 $\pm$ 1.0)$\times$10$^{6}$ & Welch \& Sage 2003 \\
\ion{H}{1} Mass (M$_{\sun}$) & (1.6 $\pm$ 0.2)$\times$10$^{6}$ & Sage \& Welch 2006 \\
\enddata
\tablenotetext{a}{Paturel \etal 2003}
\end{deluxetable}

\begin{deluxetable}{lcccccccc}
\tablewidth{0 pt}
\tablecaption{ \label{tab:ident}
  UV Magnitudes and Colors}
\tablehead{R.A. (2000) & Dec. (2000) & F330W & V-F330W & ID \\
(hr:min:sec) & ($^{\circ}$:\arcmin:\arcsec) & (mag) &	(mag)	&  }
\startdata

23:01:01.02 & +30:08:43.5  &  $22.49\pm0.03$ & $0.63\pm 0.04$ & GC11 \\
23:00:59.31 & +30:08:53.5  &  $22.53\pm0.03$ & $0.67\pm0.04$ & UV1 \\
23:01:00.82 & +30:08:47.5  &  $21.17\pm0.02$ & $0.69\pm0.03$ & L1 \\
23:00:59.56 & +30:08:38.6  &  $23.52\pm0.05$ & $0.75\pm0.07$ & UV2  \\
23:01:00.35 & +30:08:44.6  &  $23.00\pm0.04$ & $0.96\pm0.05$ & UV3 \\
23:00:59.19 & +30:08:47.5  &  $22.55\pm0.03$ & $0.98\pm0.04$ & UV4 \\
23:01:00.36 & +30:08:46.4  &  $22.59\pm0.03$ & $1.02\pm0.04$ & UV5 \\
23:01:00.24 & +30:09:01.2  &  $23.35\pm0.04$ & $1.20\pm0.05$ & UV6  \\
23:00:59.41 & +30:08:32.3  &  $22.51\pm0.03$ & $1.25\pm0.04$ & GC10 \\

\enddata
\end{deluxetable}

\newpage
\begin{deluxetable}{lccccccccc}
\tablewidth{0 pt}
\rotate
\tablecaption{ \label{tab:ident}
  Basic Data for Spectroscopic Globular Clusters in NGC 7457}
\tablehead{ID & R.A. (2000) & Dec. (2000) & $V$ & $V - I$ &
RV\tablenotemark{a}  & GC~R$_{eff}$\tablenotemark{b} & Proj.~Rad. & [m/H] & S/N \tablenotemark{c} \\
 	&	(hr:min:sec)	&	($^{\circ}$:\arcmin:\arcsec)	&	(mag)	&	(mag)	& (km s$^{-1}$)	& (pc) &(\arcsec) & (dex) & }
\startdata

GC1 & 23:00:53.0 & 30:08:44.4 & $22.78\pm0.03$ & $1.03\pm0.05$ & $860\pm15$ & 3.2 & 105 & $-1.17\pm0.18$ &  21 \\
GC2 & 23:00:53.4 & 30:08:42.3 & $22.49\pm0.03$ & $0.93\pm0.04$ & $736\pm27$ & 4.0 & 99  & $-1.30\pm0.15$ &  26 \\
GC3 & 23:00:54.2 & 30:09:23.3 & $22.30\pm0.02$ & $0.98\pm0.03$ & $908\pm20$ & 1.8 & 96  & $-1.11\pm0.13$ &  32 \\
GC4 & 23:00:56.1 & 30:09:24.2 & $22.47\pm0.03$ & $0.96\pm0.04$ & $944\pm30$ & 1.7 & 72  & $-1.33\pm0.17$ &  18 \\
GC5 & 23:00:56.7 & 30:09:26.4 & $21.97\pm0.02$ & $1.15\pm0.03$ & $918\pm18$ & 4.1 & 67  & $-0.59\pm0.06$ &  35 \\
GC6 & 23:00:57.4 & 30:08:23.3 & $22.43\pm0.03$ & $1.20\pm0.04$ & $818\pm11$ & 0.8 & 44  & $-0.28\pm0.12$ &  25 \\
GC7 & 23:00:58.0 & 30:08:33.1 & $22.34\pm0.02$ & $1.05\pm0.03$ & $822\pm10$ & 12.3 & 32 & $-0.35\pm0.18$ & 20 \\
GC8 & 23:00:58.4 & 30:09:28.1 & $22.14\pm0.02$ & $1.11\pm0.03$ & $891\pm11$ & 3.4 & 52  & $-0.94\pm0.18$ &  29 \\
GC9 & 23:00:59.2 & 30:09:21.5 & $21.76\pm0.02$ & $0.97\pm0.02$ & $941\pm32$ & 1.8 & 41  & $-1.56\pm0.07$ &  28 \\
GC10 & 23:00:59.4 & 30:08:32.3 & $21.26\pm0.01$ & $1.14\pm0.02$ & $817\pm9$ & 1.0 & 14  & $-0.67\pm0.10$ &  30 \\
N7457 & 23:01:00.0 & 30:08:42.5 &  &  & $789\pm15$ &    & 2 & $-0.28\pm0.15$ & 583 \\
GC11 & 23:01:01.0 & 30:08:43.5 & $21.86\pm0.02$ & $1.00\pm0.03$ & $721\pm19$ & 2.1 & 15 & $-0.93\pm0.06$ &  35 \\
GC12 & 23:01:01.4 & 30:09:20.6 & $21.99\pm0.02$ & $1.01\pm0.03$ & $919\pm33$ & 2.5 & 43 & $-1.11\pm0.12$ &  27 \\
GC13 & 23:01:02.1 & 30:09:26.3 & $22.84\pm0.03$ & $0.89\pm0.05$ & $834\pm27$ & 3.3 & 53 & $-1.27\pm0.09$ &  18 \\

\enddata
\tablenotetext{a}{All radial  velocities are heliocentric.}
\tablenotetext{b}{Assuming a distance of 13 Mpc}
\tablenotetext{c}{Average signal-to-noise ratio per resolution element over the bandpass of the H$\beta$ Lick/IDS index}
\end{deluxetable}

\newpage
\begin{deluxetable}{lcccccccccc}
\tablewidth{0pt}
\rotate
\tabletypesize{\footnotesize}
\tablecaption{Lick/IDS Indices
        \label{tab:indic}}
\tablehead{ID &  H$\beta$ & H$\gamma_{F}$ & H$\delta_{F}$ & CN$_{2}$ & Ca4227 & Fe4383 & Fe5270 & Fe5335 & Mg$b$ & Mg2\\
              &  (\AA) & (\AA) & (\AA) & (mag) & (\AA) & (\AA) & (\AA) & (\AA)  & (\AA) & (mag)}
\startdata

GC1 & $2.77\pm0.34$ & $0.63\pm0.34$ & $1.30\pm0.60$ & $-0.05\pm0.02$ & $1.43\pm0.33$ & $2.56\pm0.75$ & $0.95\pm0.40$ & $0.98\pm0.52$ & $0.85\pm0.38$ & $0.09\pm0.010$ \\

GC2 & $2.06\pm0.27$ & $1.67\pm0.28$ & $1.90\pm0.56$ & $0.01\pm0.02$ & $0.75\pm0.29$ & $-0.37\pm0.62$ & $1.49\pm0.38$ & $0.96\pm0.46$ & $0.03\pm0.32$ & $0.05\pm0.008$ \\

GC3 & $2.04\pm0.22$ & $0.89\pm0.26$ & $1.99\pm0.54$ & $-0.05\pm0.02$ & $0.59\pm0.28$ & $0.78\pm0.52$ & $1.91\pm0.31$ & $1.70\pm0.39$ & $1.43\pm0.25$ & $0.08\pm0.007$ \\

GC4 & $2.51\pm0.39$ & $0.79\pm0.39$ & $2.72\pm0.61$ & $-0.04\pm0.03$ & $0.63\pm0.39$ & $0.21\pm0.90$ & $1.49\pm0.50$ & $-0.62\pm0.61$ & $1.30\pm0.44$ & $0.13\pm0.012$ \\

GC5 & $1.88\pm0.20$ & $0.06\pm0.24$ & $0.53\pm0.54$  & $0.05\pm0.02$ & $1.04\pm0.27$ & $2.59\pm0.46$ & $2.03\pm0.29$ & $2.22\pm0.36$ & $2.04\pm0.23$ & $0.15\pm0.007$ \\

GC6 & $1.48\pm0.29$ & $-0.90\pm0.32$ & $0.88\pm0.59$ & $0.04\pm0.02$  & $1.62\pm0.31$ & $4.22\pm0.62$ & $2.69\pm0.37$ & $2.10\pm0.46$ & $2.75\pm0.33$ & $0.18\pm0.009$ \\

GC7 & $1.78\pm0.36$ & $-1.25\pm0.38$ & $1.11\pm0.65$ & $-0.02\pm0.02$ & $1.60\pm0.36$ & $8.07\pm0.70$ & $2.19\pm0.44$ & $2.52\pm0.52$ & $3.17\pm0.37$ & $0.21\pm0.010$ \\

GC8 & $2.22\pm0.25$  & $1.22\pm0.27$ & $1.03\pm0.56$  & $-0.07\pm0.02$ & $1.27\pm0.29$ & $3.05\pm0.54$ & $1.43\pm0.35$ & $1.34\pm0.43$ & $1.22\pm0.27$ & $0.12\pm0.008$ \\

GC9 & $2.64\pm0.25$ & $1.46\pm0.26$ & $2.54\pm0.54$  & $-0.04\pm0.02$ & $0.28\pm0.29$ & $0.59\pm0.53$ & $0.60\pm0.34$ & $0.28\pm0.44$ & $-0.17\pm0.30$ & $0.06\pm0.008$ \\

GC10 & $2.14\pm0.24$ & $-0.14\pm0.28$ & $0.52\pm0.57$  & $0.03\pm0.02$ & $0.86\pm0.30$ & $3.20\pm0.54$ & $1.51\pm0.34$ & $1.43\pm0.40$ & $2.34\pm0.26$ & $0.17\pm0.007$ \\

N7457 & $2.56\pm0.07$ & $-0.04\pm0.13$ & $1.01\pm0.48$  & $0.04\pm0.01$ & $1.22\pm0.21$ & $4.69\pm0.10$ & $2.85\pm0.16$ & $2.63\pm0.25$ & $2.94\pm0.07$ & $0.22\pm0.004$ \\

GC11 & $2.16\pm0.21$ & $1.28\pm0.22$ & $2.01\pm0.52$ & $-0.06\pm0.02$ & $0.63\pm0.26$ & $1.26\pm0.43$ & $1.68\pm0.30$ & $1.31\pm0.38$ & $0.88\pm0.24$ & $0.10\pm0.007$ \\

GC12 & $2.03\pm0.26$ & $1.42\pm0.26$ & $2.15\pm0.54$ & $-0.02\pm0.02$ & $0.38\pm0.29$ & $1.18\pm0.58$ & $1.23\pm0.37$ & $1.28\pm0.46$ & $0.77\pm0.30$ & $0.07\pm0.008$ \\

GC13 & $1.50\pm0.38$ & $0.59\pm0.36$ & $1.73\pm0.61$ & $-0.02\pm0.02$ & $0.95\pm0.36$ & $1.10\pm0.79$ & $1.18\pm0.50$ & $1.70\pm0.57$ & $0.20\pm0.41$ & $0.07\pm0.011$ \\

\enddata
\end{deluxetable}

\newpage
\begin{deluxetable}{lcccccc}
\tablewidth{0pt}
%\rotate
\tabletypesize{\footnotesize}

\tablecaption{Multi-Line Fits
        \label{tab:indic}}
\tablehead{
ID & Age & [Fe/H] & [$\alpha$/Fe] & [Z/H] & $\chi^{2}$ &\# indices\\
      & (Gyr) &         &                       &                    &                     }
\startdata

GC1 & $11\pm5$ & $-1.3\pm0.3$ & $0.4\pm0.3$ & $-1.0\pm0.2$ & $19.8$ & $15$ \\ 

GC2 & $13\pm4$ & $-1.6\pm0.3$ & $0.4\pm0.4$ & $-1.2\pm0.1$ & $17.6$ & $13$ \\

GC3 & $12\pm2$ & $-1.4\pm0.2$ & $0.3\pm0.2$ & $-1.2\pm0.1$ & $11.9$ & $15$ \\

GC4 & $11\pm4$ & $-1.6\pm0.3$ & $0.5\pm0.3$ & $-1.2\pm0.2$ & $10.0$ & $15$ \\

GC5 & $12\pm3$ & $-0.9\pm0.2$ & $0.3\pm0.2$ & $-0.6\pm0.2$ & $5.1$ & $13$ \\

GC6 & $11\pm3$ & $-0.5\pm0.1$ & $0.4\pm0.1$ & $-0.1\pm0.1$ & $5.6$ & $14$ \\

GC6\tablenotemark{**} & $15\pm3$ & $-0.4\pm0.1$ & $0.3\pm0.1$ & $-0.1\pm0.1$ & $3.2$ & $9$ \\

GC7 & $15\pm6$ & $-0.3\pm0.2$ & $0.1\pm0.2$ & $-0.2\pm0.1$ & $18.4$ & $12$ \\

GC7\tablenotemark{**} & $11\pm6$ & $-0.4\pm0.3$ & $0.2\pm0.2$ & $-0.2\pm0.2$ & $13.6$ & $8$ \\

GC8 & $7\pm3$ & $-0.8\pm0.2$ & $0.3\pm0.2$ & $-0.5\pm0.2$ & $8.0$ & $13$ \\

GC9 & $9\pm3$ & $-1.8\pm0.2$ & $0.7\pm0.2$ & $-1.12\pm0.2$ & $4.9$ & $14$ \\

GC10 & $11\pm4$ & $-0.8\pm0.2$ & $0.2\pm0.2$ & $-0.7\pm0.1$ & $7.9$ & $15$ \\

N7457 & $2.5\pm0.3$ & $0.18\pm0.06$ & $-0.03\pm0.03$ & $0.15\pm0.06$ & $7.2$ & $17$ \\

GC11 & $13\pm3$ & $-1.5\pm0.2$ & $0.3\pm0.3$ & $-1.2\pm0.1$ & $8.2$ & $15$ \\

GC12 & $12\pm3$ & $-1.5\pm0.3$ & $0.3\pm0.3$ & $-1.2\pm0.1$ & $5.0$ & $11$ \\

GC13 & $12\pm3$ & $-1.5\pm0.3$ & $0.3\pm0.3$ & $-1.2\pm0.2$ & $17.7$ & $15$ \\

\enddata
\tablenotetext{**}{All Balmer lines were excluded from these fits, due to the possibility that they are filled in by emission.}
\end{deluxetable}

\newpage
\begin{deluxetable}{lccc}
\tablewidth{0 pt}
\tablecaption{ \label{tab:eml}
  Observed Emission Lines}
\tablehead{Line & GC6 & GC7 }
\startdata

$[$\ion{O}{2}$]\lambda$3727 & Strong & None \\
$[$\ion{Ne}{3}$]\lambda$3869 & None & Weak \\
$[$\ion{Ne}{3}$]\lambda$3967 & None & Weak \\
H$\beta$ & Moderate & Very Weak \\
$[$\ion{O}{3}$]\lambda$4959 & Strong & Very Strong \\
$[$\ion{O}{3}$]\lambda$5007 & Strong & Very Strong \\

\enddata
\end{deluxetable}

\begin{deluxetable}{lccc}
\tablewidth{0 pt}
\tablecaption{ \label{tab:eml}
 Emission Line Ratios for GC6}
\tablehead{Ratio & MILES 2.0 Gyr & MILES 12.6 Gyr }
\startdata

$[$\ion{O}{2}$]\lambda$3727 / H$\beta$ & 2.9 $\pm$ 0.4 & 4.3 $\pm$ 0.9 \\
$[$\ion{O}{3}$]\lambda5007$ / H$\beta$ & 2.3 $\pm$ 0.3 & 3.9 $\pm$ 0.7 \\
$[$\ion{O}{2}$]\lambda$3727 / $[$\ion{O}{3}$]\lambda$5007 & 1.3 $\pm$ 0.1 & 1.1 $\pm$ 0.1 \\

\enddata
\end{deluxetable}

\end{document}